\documentclass[twoside,11pt]{article}

\usepackage{blindtext}

\usepackage[abbrvbib, preprint]{jmlr2e}

\usepackage{bm}
\usepackage{bbm}
\usepackage{mathtools}
\usepackage{tikz}
\usepackage{multirow,array}
\usepackage[capitalise]{cleveref}
\theoremstyle{plain}
 
\newtheorem{theorem}{Theorem}
 
\newtheorem{proposition}[theorem]{Proposition}

\theoremstyle{definition}
\newtheorem{definition}[theorem]{Definition}

\usepackage{soul}

\usetikzlibrary{shapes,decorations,arrows,calc,arrows.meta,fit,positioning}
\usetikzlibrary{positioning,angles,quotes,intersections}
\usetikzlibrary{through}
\tikzset{
    -Latex,auto,node distance =1 cm and 1 cm,semithick,
    state/.style ={circle, draw, minimum width = 0.7 cm},
    point/.style = {circle, draw, inner sep=0.04cm,fill,node contents={}},
    bidirected/.style={Latex-Latex,dashed},
    el/.style = {inner sep=2pt, align=left, sloped},
    edge/.style = {->},
    da/.style ={ellipse, draw, dashed, minimum width = 0.4 cm},
}

\usetikzlibrary{arrows,shapes.arrows,shapes.geometric,shapes.multipart,
decorations.pathmorphing,positioning,}

\usetikzlibrary{graphs,graphs.standard}
\usepackage{float}

\newcommand{\Pa}{\mathrm{pa}}

\newcommand{\p}{\textbf{p}}
\DeclareMathOperator{\Do}{do}
\newcommand{\Prob}{\mathbb{P}}
\newcommand{\widebar}{\overline}
\newcommand{\R}{\mathbb{R}}

\newcommand{\Xir}{X_{\mathrm{i}}}
\newcommand{\Xor}{X_{\mathrm{o}}}
\newcommand{\Yir}{Y_{\mathrm{i}}}
\newcommand{\Yor}{Y_{\mathrm{o}}}
\newcommand{\xir}{x_{\mathrm{i}}}
\newcommand{\xor}{x_{\mathrm{o}}}
\newcommand{\yir}{y_{\mathrm{i}}}
\newcommand{\yor}{y_{\mathrm{o}}}
\newcommand{\ir}{\mathrm{i}}
\newcommand{\orr}{\mathrm{o}}

\usepackage{lastpage}
\jmlrheading{}{2025}{1-\pageref{LastPage}}{}{}{}{M.~Scauda, J.~Kuipers, and G.~Moffa}

\ShortHeadings{A Latent Causal Inference Framework for Ordinal Variables}{M.~Scauda, J.~Kuipers, and G.~Moffa}
\firstpageno{1}
\allowdisplaybreaks
\begin{document}

\title{A Latent Causal Inference Framework for Ordinal Variables}

\author{\name Martina Scauda \email ms2985@cam.ac.uk \\
       \addr Dep.\ of Mathematics and Computer Science, University of Basel, Basel, Switzerland\\
       Dep.\ of Pure Mathematics and Mathematical Statistics, University of Cambridge, Cambridge, UK
       \AND
       \name Jack Kuipers \email jack.kuipers@bsse.ethz.ch \\
       \addr Dep.\ of Biosystems Science and Engineering,
        ETH Zurich, Basel, Switzerland
        \AND 
        \name Giusi Moffa \email giusi.moffa@unibas.ch  \\
        \addr Dep.\ of Mathematics and Computer Science, University of Basel, Basel, Switzerland
}

\editor{My editor}

\maketitle

\begin{abstract}%   <- trailing '%' for backward compatibility of .sty file
Ordinal variables, such as on the Likert scale, are common in applied research. Yet, existing methods for causal inference tend to target nominal or continuous data. When applied to ordinal data, this fails to account for the inherent ordering or imposes well-defined relative magnitudes. Hence, there is a need for specialised methods to compute interventional effects between ordinal variables while accounting for their ordinality. One potential framework is to presume a latent Gaussian Directed Acyclic Graph (DAG) model: that the ordinal variables originate from marginally discretising a set of Gaussian variables whose latent covariance matrix is constrained to satisfy the conditional independencies inherent in a DAG. Conditioned on a given latent covariance matrix and discretisation thresholds, we derive a closed-form function for ordinal causal effects in terms of interventional distributions in the latent space. Our causal estimation combines naturally with algorithms to learn the latent DAG and its parameters, like the Ordinal Structural EM algorithm. Simulations demonstrate the applicability of the proposed approach in estimating ordinal causal effects both for known and unknown structures of the latent graph. As an illustration of a real-world use case, the method is applied to survey data of 408 patients from a study on the functional relationships between symptoms of obsessive-compulsive disorder and depression.
\end{abstract}


\section{Introduction}
Ordinal or ordered categorical variables, which take categorical values following an intrinsic listing order, are common in many research fields \citep{agrestiAnalysisOrdinalCategorical2010}. Examples include letter grades (A, B, C, D, E, F), survey questions on a Likert scale (strongly disagree, disagree, undecided, agree,
strongly agree), stages of cancer (I, II, III, IV); as well as discretised continuous data, such as age groups (children, youth, adults, seniors). The latent continuous underlying construct may be unavailable or non-observable for practical or confidentiality reasons. In such a context, for analysing their ordinal values it is natural to conceptualise them as obtained by marginally discretising a set of latent continuous variables. 

Given the ubiquity of ordinal data, developing interpretable and robust methodologies for defining and inferring the causal effects of interventions on ordinal outcomes is highly relevant. Because of intrinsic technical difficulties, much of the causal inference literature treats ordinal data as nominal or continuous, either ignoring the inherent order among the categories or assuming that the scale between ordinal levels corresponds to well-defined relative magnitudes. 

Under the potential outcomes framework \citep{neyman, Rubin1974EstimatingCE}, causal effects are usually defined as comparisons between the potential outcomes under the two levels of a binary treatment. For an ordinal response, the most common parameter of interest, the average causal effect, is not well-defined since a notion of average only exists for continuous and binary outcomes, where the average is the proportion. Analogously, other measures of centrality do not work with an ordinal response because the scale of the categories is not well-defined.  For these reasons, most of the literature has focused on binary outcomes \citep{rosrub} as a special case of ordinal ones, with some exceptions. Notable examples include \citet{Rosenbaum2001}, who explored causal inference for ordinal outcomes under the assumption of monotonicity, stating that the treatment has a non-negative effects for all units; \citet{agrestiAnalysisOrdinalCategorical2010} and \citet{agresti_ordinal_2017} who examined a range of ordinal effect measures while assuming independence between potential outcomes; and  \citet{volfovsky_causal_2015} who adopted a Bayesian approach that required specifying a full parametric model for the joint distribution of potential outcomes. In the setting of randomised trials, \citet{diaz2016} proposed a robust causal parameter that avoided reliance on the proportional odds model assumption.  Another challenge arises with ordinal treatments since, in the context of the potential outcomes framework, the estimation is often performed using propensity score methods, which are generally confined to binary treatment scenarios. For univariate ordinal treatment variable, \citet{Joffe} proposed an extension of the propensity score method and \citet{imai_causal_2004} generalised it to arbitrary treatment regimes (including ordinal and even multivariate) by using propensity functions. As an alternative, \citet{imbens_role_2000} adopted inverse probability weighting for estimating causal effects from ordinal treatments.

Alongside the potential outcomes framework, causal diagrams, represented by directed acyclic graphs (DAGs) provide a complementary approach to causal inference. Using causal diagrams to describe a data-generating mechanism is appealing as the edges may naturally encode causal links between the variables represented by the nodes on the graph. For a known causal DAG, the intervention calculus \citep{pearl_2009} provides a machinery to determine the causal effects of one variable on another. In more common practical situations a suitable causal diagram is rarely known, so we need methods that can learn both a network structure and its parameters from data, keeping in mind that observational data only identify the DAG describing their joint distribution up to its Markov equivalence class \citep{vermapearl1990}, which we may represent as a completed partially directed acyclic graph (CPDAG).  

Learning Bayesian networks intrinsically depends on the data type with existing solutions mainly focused on continuous and nominal data \citep{rios2021benchpress, kitson2023survey}. 
In the case of Gaussian data, to overcome the lack of knowledge about the graphical structure \citet{maathuis_estimating_2009} provided lower bounds for the causal effects after finding a Markov equivalence class compatible with the data. By adopting a Bayesian approach, one can integrate structure learning and effect estimation into a procedure that yields the posterior distribution of causal effects. A key advantage is that the method captures both the graphical and parameter uncertainty, as originally proposed and illustrated in a psychology application by \citet{moffa2017psychoBayesian} for binary data. Later works extended this Bayesian procedure to dynamic Bayesian networks \citep{kuipers2019psychodbn} and linear Gaussian models \citep{Viinikka2020}.
 
Until recently, the literature on causal graphical models has given little consideration to the problem of determining a suitable causal graphical framework for ordinal data while coherently defining and evaluating ordinal causal effects. For the structure learning part, \citet{luo_learning_2021} designed an Ordinal Structural EM (OSEM) algorithm accounting for the ordinality through a \emph{latent Gaussian DAG-model}. The parametrization relies on a multivariate probit model, where each ordinal variable is the marginal discretisation of a latent Gaussian variable, with their joint Gaussian distribution factorising according to a DAG.
In a similar vein, \citet{castelletti_bayesian_2021} assumed a DAG-probit model to evaluate intervention effects in a Bayesian framework, in the special case when the intervention variables are continuous and the response is binary. They model the binary outcome as a discretised instance of a continuous underlying variable, whose distribution, jointly with the remaining continuous variables, is Gaussian and obeys the conditional independence relationships inherent in a DAG. Since binary variables are a special case of ordinal variables, the more general latent Gaussian DAG-model adopted in \citet{luo_learning_2021} encompasses the \emph{Probit DAG-model} considered in \citet{castelletti_bayesian_2021}.

Recently \cite{grzegorczyk_being_2024} proposed a Bayesian version of the OSEM algorithm \citep{luo_learning_2021}, while \citet{castelletti2023learning} extended the structural recovery approach of \citet{castelletti_bayesian_2021} to deal with mixed data, including ordinal. Nevertheless, the causal graphical framework still lacks a methodology to determine the causal effect on an ordinal outcome following an intervention on another binary or potentially ordinal variable. 

Realising the theoretical gap in this matter and building on the multivariate probit regression models of \citet{albeertchib} and the latent Gaussian modelling adopted in \cite{luo_learning_2021} for structure learning, we develop an order-preserving methodology to compute interventional effects among ordinal variables in a latent Gaussian DAG model. 
The rest of the article is organised as follows. In \cref{sec2}, we provide an overview of previous approaches including both potential outcomes and causal diagrams, along with latent Gaussian DAG modelling. In \cref{sec3}, we present our original contribution with the definition of ordinal causal effects, and how to evaluate them in the proposed latent Gaussian DAG framework. In \cref{sec4}, we use synthetic and real data from \citet{McNally} to illustrate the performance of our proposed methodology in terms of causal effect estimation with known and unknown latent DAG structures. Lastly, in \cref{sec5}, we discuss our results in relation to previous approaches and highlight possible extensions of the present work.

\section{Background }\label{sec2}
Let $\boldsymbol{X} = (X_1, \dots, X_n)^\intercal$ be a collection of $n$ ordinal variables, where $X_m$ takes values in the set $\{\tau(m,1), \dots, \tau(m,L_m)\}$ with $\tau(m,1)< \dots < \tau(m,L_m), \, m=1, \dots, n.$ Each variable is assumed to be at least binary, meaning the number of levels $L_m \geq 2$. It is conventional to set $\tau(m,j)= j-1$ for all $ 1 \leq j \leq L_m$. 

We are interested in the causal effect on an outcome variable $\Xor$ of a deterministic intervention on an intervention variable $\Xir$. To describe intervention effects we use Pearl's do-operator \citep{pearl_2009} where the distribution of $\Xor$ under an intervention on $\Xir$ is generally denoted as $\Prob(\Xor = \xor \mid \Do(\Xir = \xir))$.
Distributional changes or distribution shifts in the outcome variable between different levels of the intervention variable often constitute target estimands of practical relevance \citep{holland_causal_1988}. Evaluating and comparing the change in the probability of $\Xor$ belonging to level $k$, when setting the intervention variable $\Xir$ to level $l'$ vs level $l$ provides a measure of the distribution shift
\begin{equation}\label{target}
       \Prob\Bigl[\Xor = \tau(\orr,k) \mid \Do\bigl(\Xir = \tau(\ir,l')\bigr)\Bigr] - \Prob\Bigl[\Xor = \tau(\orr,k) \mid \Do\bigl(\Xir = \tau(\ir,l)\bigr)\Bigl] 
\end{equation}
for each $l \neq l'$ and $l,l'\in \{1,\dots, L_\ir\}$ and $k\in \{1,\dots, L_\orr\}$. Our objective is to evaluate \textit{ordinal causal effects} (OCEs) as represented by the target causal estimands in \cref{target}. 

\subsection{Related Work}
Consider a study with $N$ units, an ordinal outcome $X$ with $K$ observed levels and a binary treatment. Under the stable unit treatment value assumption (SUTVA) \citep{Imbens_Rubin_2015}, we can define the pair of potential outcomes $\{X_m(1), X_m(0)\}$ of unit $m$ under treatment and control. In this framework, we can consider causal estimands for ordinal outcomes on either the observed scale or the latent scale. In the latter, there is a pair of latent potential outcomes $(Z_m(1), Z_m(0))$ for each unit $m$ and a determinist function $g(\cdot)$ which maps them into the observed scale as $X_m(1)=g(Z_m(1))$ and $X_m(0)=g(Z_m(0))$. If the map $g$ is known explicitly or fully identified, the continuous latent potential outcomes can be treated as the actual outcomes and causal analysis reduces to the classical results for continuous potential outcomes \citep{Imbens_Rubin_2015}. In practice, $g$ is often inferred from data and likely its location and scale lack identifiability, key features needed to define meaningful estimands on the latent scale. Nonetheless, \citet{volfovsky_causal_2015} adopts a latent variable formulation to determine posterior predictive estimates of the induced estimands on the observed scales. 

On the observed scale instead, the joint distribution of the potential outcomes, which contains complete information about any causal effect, can be summarized by a matrix $\bm{P}$ with entries $p_{kl}= \Prob(X(0) = k, X(1) = l) \quad k,l=0, \dots, K-1$, that are the proportions of units whose potential outcomes take on those categories.  All estimands are therefore functions of the matrix $\bm{P}$. Since the relative magnitudes between the pair of categories are not defined for ordinal data, a meaningful one-dimensional summary should summarise the differences between the marginal distributions of the potential outcomes. However, this is often insufficient to characterize causal effects and multidimensional estimands are required, such as distributional causal effects \citep{Ju2010CriteriaFS}
\begin{equation}\label{eqdiseffects}
   \Delta_j = \Prob[X_m(1) \geq j] - \Prob[X_m(0) \geq j] = \sum_{k \geq j} \sum_{l=0}^{K-1} p_{kl} - \sum_{l \geq j} \sum_{k=0}^{K-1} p_{kl},
\end{equation}
which measure the difference between the marginal distributions of the potential outcomes at different levels $j$, similarly to the effects discussed in \citet{Boes2013NonparametricAO}. Nonetheless, unless those effects have the same sign for all $j$, it is difficult to select the preferable treatment. 

Another possible solution to measure the magnitude of the effect relative to treatment is to use causal estimands that conditions on the level of the potential outcome under control, such as $K$-dimensional summaries of the form $M_{1i}= \text{median}[X(1) \mid X(0) = j]$  or $M_{2i}= \text{mode}[X(1) \mid X(0) = j] \quad (j=0, \dots, K-1)$ \citep{volfovsky_causal_2015}.   
%which depend on the joint distribution of potential outcomes and hence required modelling assumptions about the potential outcomes. 
As underlined by \citet{lupeng}, neither of the previous estimands is a direct measure of the treatment effect. Moreover, the conditional medians may not be unique and they are only well-defined for the levels $j$ such that the sum of the $j$-th column of $\bm{P}$ is positive.

Alternatively, \citet{lupeng} propose the following two causal estimands for ordinal outcomes, which measure the probabilities that the treatment is beneficial and strictly beneficial for the experimental units respectively 
\begin{equation}
    \iota = \Prob[X_m(1) \geq X_m(0)]= \underset{k \geq l}{\sum \sum} \, p_{kl},  \quad \eta = \Prob[X_m(1) > X_m(0)]= \underset{k > l}{\sum \sum} \, p_{kl}
\end{equation}
Alternatively, \citet{agresti_ordinal_2017} used the relative treatment effect for ordinal outcomes defined as 
\begin{equation}
    \gamma =  \Prob[X_m(1) > X_m(0)] - \Prob[X_m(1) < X_m(0)] = \iota + \eta -1
\end{equation}
When $K=2$ (i.e.\ $X$ is binary), the relative treatment effect reduces to
\begin{equation}\gamma = \mathbb{E}[X_m(1)-X_m(0)]= \mathbb{E}[X_m(1)] - \mathbb{E}[X_m(0)]
\end{equation}
which is exactly the average treatment effect. 
Since these estimands depend on the association between the two potential outcomes, they are not identifiable from observed data. Under the partial identification strategy of \citet{Manski}, \citet{lupeng} derive sharp bounds on $\iota$ and $\eta$ in closed form. \citet{lu_sharp_2020} extend these bounds to $\gamma$, while \citet{chiba_bayesian_2018} propose a Bayesian approach requiring a prior on the joint distribution of potential outcomes. 

Estimating causal effects for ordinal variables in a latent Gaussian DAG model is closely related to estimating effects under the assumption of Gaussianity. Therefore we start by defining graphical models and causal effects in a Gaussian set-up and build on it for defining and evaluating effects with ordinal data. 
To fix the notation we start by defining graphical models and briefly introducing general intervention calculus, closely following \citet{pearl_2009}.

\subsection{Graphical Models and Causal Effects}\label{subCausEffect}
Let $\mathcal{G} = (V, E)$ be a DAG, with $V = \{1, \dots, n\}$ denoting the finite set of vertices (or nodes) and $E\subset V \times V$ the set of directed edges, whose elements are $(h,j) \equiv h \rightarrow j$ such that $(h,j) \in E$ but $(j,h) \notin E$. Moreover, $\mathcal{G}$ contains no cycles, i.e.\ paths of the form $k_0, k_1, \dots, k_m$ such that $k_0 \equiv k_m$. For a given vertex $j$, node $h$ is called a parent of $j$ and node $j$ a child of $h$ if $(h,j) \in E$. The parent set of a node $j$ is denoted by $\Pa(j)$. 

We consider a collection of $n$ random variables $\bm{Y}=(Y_1, \dots, Y_n)^{\intercal}$ and assume that their joint p.d.f.\ $f(\bm{y})$, fully characterised by a set of parameters $\theta$, is Markovian with respect to the DAG $\mathcal{G}$, meaning it factorises as 
\begin{equation}\label{facdistribution}
    f(y_1, \dots, y_n \mid \theta,\mathcal{G} ) = \prod_{m=1}^n f(y_m \mid y_{\Pa(m)}, \theta_m,\mathcal{G})
\end{equation}
where $\bm{y}$ is a realisation of $\bm{Y}$, $\theta = \cup_{m=1}^n \theta_m$ and the subsets $\{\theta_m\}_{m=1}^n$ are assumed to be disjoint. The pair $\mathcal{B} = (\mathcal{G}, \theta)$ denotes a Bayesian network, a specific family of probabilistic graphical model where the underlying structure is a DAG 
\citep[see Section 3.2.2 of][]{lauritzen1996graphical}.  
From this point forward, all arguments implicitly condition on a given DAG $\mathcal{G}$, without explicitly indicating it in the notation.

The joint distribution $f(\bm{y})$ constitutes the observational (or pre-intervention) distribution.
To denote a deterministic intervention on a variable $\Yir,  \ir \in \{1, \dots, n\}$, \citet{pearl_2009} introduced the do-operator $\Do( \Yir = \yir')$, which consists in enforcing $\Yir = \yir'$ uniformly over the population.  For a Markovian model, the post-intervention density is then given by the following truncated factorisation formula: 
\begin{equation}\label{truncformula}
    f(y_1, y_2, \dots, y_n \mid \Do(\Yir = \yir')) = \begin{cases}
    \prod_{j \neq \ir} f(y_j \mid y_{\Pa(j)}) \quad &\text{if $\yir = \yir'$} \\
    0 \quad &\text{if $\yir \neq \yir'$}  \\
    \end{cases}
\end{equation} 
where $f(y_j \mid y_{\Pa(j)})$ are the pre-intervention conditional distributions. 

The post-intervention distribution of a variable $Y_j$ with $j \neq \mathrm{i}$ is then obtained by integrating out $y_1, \dots, y_{j-1}, y_{j+1} \dots, y_n$ in \cref{truncformula}, simplifying to 
\begin{equation}\label{simpletruncformula}
    f(y_j \mid \Do(\Yir = \yir')) = \begin{cases}
    f(y_j) &\text{if $Y_j \in y_{\Pa(\ir)}$} \\
    \int f(y_j \mid \yir', y_{\Pa(\ir)}) f(y_{\Pa(\ir)}) \text{d}y_{\Pa(\ir)} \quad &\text{if $Y_j \notin y_{\Pa(\ir)}$}  \\
    \end{cases}
\end{equation}
where $f(\cdot)$ and $f(\cdot \mid \yir', y_{\Pa(\ir)})$ represent pre-intervention distributions \citep[][page 73]{pearl_2009}. The previous expression for $Y_j \notin y_{\Pa(\ir)}$ is a special case of the back-door adjustment \citep[][page 79]{pearl_2009} since $\Pa(\ir)$ satisfies the back-door criterion on $(\Yir, Y_j)$ when $Y_j \notin y_{\Pa(\ir)}$.

\subsection{Gaussian DAG-models}
If the joint distribution of $\bm{Y}$ is Gaussian, we have 
\begin{equation}
\bm{Y} \sim \mathcal{N}(\bm{\mu}, \bm{\Sigma}), \end{equation}
where the precision matrix $\bm{\Omega} = \bm{\Sigma}^{-1}$ is symmetric, positive definite and Markov relative to $\mathcal{G}$.
By the normality assumption, the Gaussian DAG-model is guaranteed to be faithful to the DAG almost everywhere in the parameter space, implying that the conditional independence relationships entailed by the distribution are exactly the same as those encoded by the DAG via the Markov property \citep[p.~48]{pearl_2009}. 
For a Gaussian DAG-model, the factorisation in \cref{facdistribution} takes the form (see e.g.~\cite{geiger2002})
\begin{equation}\label{facnormal}
    f(y_1, \dots, y_n \mid \mathcal{G}, \bm{\mu}, \bm{\Sigma}) = \prod_{m=1}^n \phi(y_m \mid \mu_m(y_{\Pa(m)}), \sigma_m^2),
\end{equation}
where $\phi$ denotes the univariate normal density function. In this case, \cref{facdistribution} can also be written as a linear structural equation model (SEM) 
\begin{equation}\label{linear SEM}
    Y_j = \beta_{0j} + \sum_{h \in \Pa(Y_j)} \beta_{hj}Y_h + \epsilon_j, \quad \forall \,j=1, \dots, n,
\end{equation}
where the parameters $\beta_{hj}$ are called the \emph{path coefficients} and the error terms $\epsilon_1, \dots, \epsilon_n$ are mutually independent with mean 0 and finite variance $\bm{V} = \text{diag}(\bm{v})$, with $\bm{v}$ the $(n,1)$ vector of variances whose $m$-th element is $v_m = \text{Var}(\epsilon_m)$. 
If we assume a topological ordering of the vertices, meaning there exists a label permutation $(\pi_1, \dots, \pi_n)$ of the vertices $V=(1, \dots, n)$ such that $(h,j) \in E$ implies $\pi_h < \pi_j$, the matrix $(\bm{B})= \{\beta_{jh}, h \leq j\}$, the transpose of the matrix of path coefficients, is lower-triangular and the covariance matrix $\bm{\Sigma}$ is symmetric and positive definite, with the following Cholesky decomposition \citep{silva_hidden_2009}  
\begin{equation}\label{choldec}
    \bm{\Sigma} = (\bm{I} - \bm{B})^{-1} \bm{V} (\bm{I} - \bm{B})^{-\intercal}. 
\end{equation}
By induction on the number of vertices $V$, one can show that such topological ordering always exists, even if it is not unique in general. 

\cref{linear SEM} can be fitted using linear regression and is called structural because it is assumed to hold for the interventional distribution as well with the exception of the intervened upon variable. Given a linear SEM, the total causal effect $\beta(Y_h \to Y_j)$, of $Y_h$ on $Y_j$, may be evaluated as the product of the path coefficients along all directed path (causal paths) $\mathcal{C}(h,j)$ from vertex $h$ to $j$ \citep{wright}.
\begin{equation}\label{eqeffectsem}
    \beta(Y_h \to Y_j) =  \sum_{(k_0, \dots, k_m) \in \mathcal{C}(h,j)} \prod_{l=1}^m \beta_{k_{l-1} k_l} = \left[(\bm{I} - \bm{B})^{-\intercal}\right]_{hj}
\end{equation}
It is common to summarize the post-intervention distribution of \cref{simpletruncformula} by its mean $\mathbb{E}(Y_j \mid \Do(\Yir = \yir')$ and when $Y_j$ is continuous, the total causal effect of $\Do(\Yir = \yir')$ on $Y_j$ is defined as 

\begin{equation}\label{dercausal}
    \frac{\partial}{\partial y_\mathrm{i}} \mathbbm{E}(Y_j \mid \Do(\Yir = y_\mathrm{i})) \Big |_{y_\mathrm{i} = \yir'}.
\end{equation}
Due to the linearity of expectation in the Gaussian case, the mean under the post-intervention distribution in \cref{simpletruncformula} becomes \begin{equation}\label{regcausal}
\mathbbm{E}(Y_j \mid \yir', y_{\Pa(\ir)})=\lambda_{0j} + \lambda_{\mathrm{i}} \yir' + \sum_{h \in \Pa(Y_j)} \lambda_{hj}Y_h.\end{equation}
One can show that in this scenario the causal effect of $\Yir$ on $Y_j$  with $Y_j \notin \Pa(\Yir)$ corresponds to the regression parameter $\lambda_\mathrm{i}$ associated to the variable $\Yir$ in \cref{regcausal} \citep[see p.~3138 of ][] {maathuis_estimating_2009}. For this reason, in the Gaussian case, the causal effect does not depend on the intervention value $\yir'$ and can be interpreted for any value of $\yir'$ as \[\mathbbm{E}(Y_j \mid \Do(\Yir = \yir' +1)) - \mathbbm{E}(Y_j \mid \Do(\Yir = \yir')). \]

To model a binary response $X_1$ potentially affected by a set of observed continuous variables $(Y_2, \dots, Y_n)$ , \citet{castelletti_bayesian_2021} assumed a DAG-Probit model, where the observable binary outcome is obtained by discretising a continuous latent variable $Y_1$ 
\begin{equation}
    X_1 = \begin{cases}
        1 \quad &\text{if} \, Y_1 \in  [\alpha_1, + \infty)\\
        0 \quad &\text{if} \, Y_1 \in  (- \infty, \alpha_1)\\
    \end{cases} \quad \text{for}  \, \alpha_1 \in (-\infty, +\infty)
\end{equation}
and the joint distribution of $(Y_1, \dots,Y_n)$ is Gaussian and Markov w.r.t.\ $\mathcal{G}$ as in \cref{facnormal}. In their construction the latent outcome variable is the last one in topological order and therefore it cannot have children. 

Under this setting, to determine the effect of an intervention on the observable response variable, one may evaluate  
\begin{equation}\label{expcast}
\begin{split}
    \mathbbm{E}(X_1 \mid \Do(\Yir = \yir'), \bm{\mu}, \bm{\Sigma}, \alpha_1, \mathcal{G}) 
    &= \Prob(X_1 = 1 \mid \Do(\Yir = \yir'), \bm{\mu}, \bm{\Sigma}, \alpha_1, \mathcal{G}) \\
    &= \Prob(Y_1 \geq \alpha_1 \mid \Do(\Yir = \yir'), \bm{\mu}, \bm{\Sigma}, \mathcal{G}) \\
    &= 1- \Phi\Big(\frac{\alpha_1 - \mu_1 - \gamma_{\mathrm{i}}\yir'}{\xi_1^2}\Big),
\end{split} 
\end{equation}
  where both $\gamma_{\mathrm{i}}$ and $\xi_1^2$ can be expressed in terms of product between submatrices of $\bm{\Sigma}$, as they are determined by the specialization of the post-interventional distribution in  \cref{simpletruncformula} to the case of the latent Gaussian $Y_1$
\citep[see Prop. 3.1. of][]{castelletti_bayesian_2021}.

Similarly to \cref{dercausal}, one may compute
$$\frac{\partial}{\partial \yir} \mathbbm{E}(X_1 \mid \Do(\Yir = \yir), \bm{\Sigma}, \theta_0, \mathcal{G}) \Big |_{\yir = \yir'},$$ 
but this will still depend on $\yir'$ (unlike in the Gaussian continuous case), and the parameters of the DAG Probit model. For this reason, and because \cref{expcast} admits an intuitive interpretation as a probability, \citet{castelletti_bayesian_2021} simply denote $\Prob(X_1 = 1 \mid \Do(\Yir = \yir'), \bm{\mu}, \bm{\Sigma}, \alpha_1, \mathcal{G})$ as the causal effect on $X_1$ due to an intervention $\Do(\Yir = \yir')$.

\subsection{Latent Gaussian DAG-model}\label{seclatentmodel}
To model ordinal variables, we assume that the variables in the Gaussian vector $\bm{Y} = (Y_1, \dots, Y_n)^{\intercal}$ are unobserved, and we observe instead ordinal variables obtained from the continuous variables by discretisation. The diagram in \cref{beyondtoyDAG} provides a visual representation of the set-up in an illustrative case with a handful of variables.   
Each ordinal variable $X_m$ is assumed to be a discretised version of the latent variable $Y_m$. Specifically, given a vector of thresholds $\bm{\alpha}_m = (-\infty \coloneqq \alpha(m,0), \dots, \alpha(m, L_m)\coloneqq \infty)^{\intercal}$
\begin{equation}\label{discretization}
    X_m = \begin{cases}
        \tau(m, 1) & \text{if} \; Y_m\in (-\infty, \alpha(m,1)) \\
        \quad \vdots \\
        \tau(m, L_m) & \text{if} \; Y_m \in [\alpha(m,L_m-1), + \infty).
    \end{cases}
\end{equation}

Formally, the latent (Gaussian) DAG-model for ordinal variables is defined by
\begin{equation}\label{LATENTDAGEq}
\begin{split}
& Y_m \mid \boldsymbol{y}_{\Pa(m)}, \mathcal{\vartheta}_m, \mathcal{G} \sim \mathcal{N}\Big(\mu_m + \sum_{j \in \Pa(m)} b_{jm}(y_j-\mu_j), v_m\Big) \\
& \mathbb{P}(X_m = \tau(m,l)\mid Y_m = y_m, \boldsymbol{\alpha}_m) = \mathbbm{1}\Big(y_m \in [\alpha(m,l-1), \alpha(m,l)]\Big), \quad l=1, \dots, L_m \\
& p(\boldsymbol{x}, \boldsymbol{y} \mid \theta, \mathcal{G}) = \prod_{m=1} ^n \phi(y_m \mid \boldsymbol{y}_{\Pa(m)}, \vartheta_m, \mathcal{G})p(x_m \mid y_m, \boldsymbol{\alpha}_m) 
\end{split}
\end{equation}
where $\theta = \cup_{m=1}^n \theta_m$ with $\theta_m = (\vartheta_m, \boldsymbol{\alpha}_m), \; \vartheta_m = (\mu_m, \boldsymbol {b}_m, v_m)$ and $\boldsymbol{b}_m=(b_{jm})_{j \in \Pa(m)}$ for all $m=1, \dots, n.$  

To ensure model identifiability, we require additional constraints. In fact, different hidden Gaussian variables $\bm{Y}$ might generate the same contingency table for the ordinal variables $\bm{X}$, by shifting and scaling the thresholds according to the corresponding means and variances.

Computationally, the most convenient constraint consists in standardizing each latent dimension, as adopted by \citet{luo_learning_2021}, and consists in setting $\mu_m = 0$ for all $m=1, \dots, n$, $\bm{V} = \bm{I}$ and replacing the covariance matrix $\bm{\Sigma}$ with its correlation form $D^{-1}\bm{\Sigma} D^{-1}$, where $D$ is a diagonal matrix with elements $d_m =\sqrt{\bm{\Sigma}_{mm}}$. Nonetheless, in the next section,  we derive the causal effects for a generic covariance matrix $\bm{\Sigma}$ and mean vector $\bm{\mu}$ to ensure that the framework remains applicable to the most general case.

\section{Causal Effects in the Latent Gaussian DAG Model}\label{sec3}

\begin{figure}[t!]
    \centering
    \begin{tikzpicture}[scale=0.2,baseline=(current bounding box.north), state/.style={circle, draw=blue!30!black, fill=blue!30, minimum size=2em}]
    \node[state] (m) at (0,0) {$Y_1$};
    \node[state] (i) [above left =of m, shift={(-.5,0)}] {$\Yir$};
    \node[state] (h) [right =of m, shift={(.5,0)}] {$Y_3$};
    \node[state] (j) [right =of h] {$\Yor$};
    \node[state] (n) [above left =of j] {$Y_4$};   
    \node[da, fill=blue!60!green!10] (k) [below right =of m] {$X_1$};
    \node[da, fill=blue!60!green!10] (v) [below right =of j] {$\Xor$};
    \node[da, fill=blue!60!green!10] (b) [below right = of h] {$X_3$};
    \node[da, fill=blue!60!green!10] (c) [right = of i] {$\Xir$};
    \node[da, fill=blue!60!green!10] (d) [right = of n] {$X_4$};
    \path (m) edge (h);
    \path (i) edge (h);
    \path (i) edge (m);
    \path (h) edge (j);
    \path (n) edge (j);
    \draw[dashed, draw=blue!30!black, ->] (m) -- (k);
    \draw[dashed, draw=blue!30!black, ->] (j) -- (v);
    \draw[dashed, draw=blue!30!black, ->] (i) -- (c);
    \draw[dashed, draw=blue!30!black, ->] (n) -- (d);
    \draw[dashed, draw=blue!30!black, ->] (h) -- (b);
  \end{tikzpicture} 
  \caption{Example of latent Gaussian five-nodes DAG. Variables $X_m, \, m=1, \dots 5$ are ordinal, each obtained by discretising a latent variable $Y_m$ with associated Gaussian parameters $\theta_m$. Ordinal nodes are dashed for clarity.}
  \label{beyondtoyDAG}
  \end{figure}
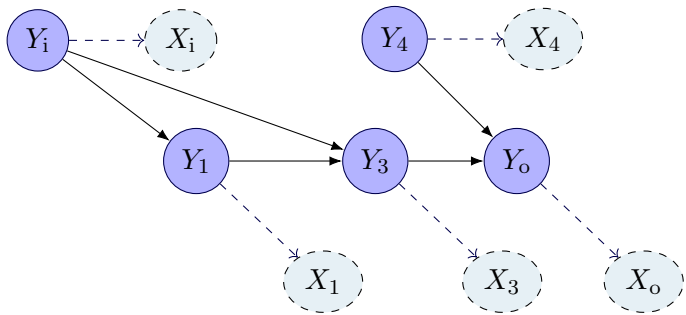  

Consider the general latent Gaussian DAG-model of \cref{seclatentmodel} and assume that the parameters $\theta$ are given together with the DAG $\mathcal{G}$.
Our interest is in computing the target causal estimand in \cref{target}, representing the OCE on $\Xor$ (outcome variable) of an intervention on  $\Xir$ (intervention variable). Since there is no causal path in the DAG $\mathcal{G}$ between the ordinal $\Xir$ and $\Xor$ (as in the illustrative graphical representation in \cref{beyondtoyDAG}), the direct computation of \cref{target} would lead to 0 for each level of the intervention and outcome variable. 

Despite the absence of a causal path between the ordinal variables $\Xir$ and $\Xor$, it is evident that they are causally related to each other through their corresponding latent parent variables. Hence, we can consider that if we were to intervene on the latent variable $\Yir$ in a way that changes the level of its ordinal child variable $\Xir$ and then compute the effect of this intervention on the latent parent $\Yor$ of $\Xor$, it is possible that the level assumed by $\Xor$ would also change as a consequence. The potential change in the level of $\Xor$ determined by an intervention on the latent parent of $\Xir$ that shifts its level then becomes the effect of interest here. Using the discretising thresholds  $\bm{\alpha}=\cup_{m=1}^n \bm{\alpha}_m$, the target causal estimand in \cref{target} on the ordinal variables can be equivalently computed as the following difference in probabilities on the latent scale. 

\begin{definition}[Ordinal Causal Effect]\label{OCEDef}
Let $\bm{Y} =(Y_1, \dots, Y_n) \sim \mathcal{N}(\bm{0}, \bm{\Sigma})$ be the underlying vector of variables in a Latent Gaussian DAG model. Let $\Yir, \, \ir \in \{1, \dots, n\}$ be the latent intervention variable and $\Yor, \, \orr \in \{1, \dots, n\} \setminus \{\ir\}$ the latent outcome variable. The Ordinal Causal Effect (OCE) on $\Yor$ of intervening on $\Yir$ is given by 
\begin{equation}\label{OCE}
\begin{split}
     \text{OCE}_{\ir \orr}(k, l \to l') &= \Prob\Big[\Yor \in [\alpha(\orr,k-1),\alpha(\orr,k)]\mid \Do(\Yir\in [\alpha(\ir,l'-1),\alpha(\ir,l')])\Big]\\
        & \quad - \Prob\Big[\Yor \in [\alpha(\orr,k-1),\alpha(\orr,k)] \mid \Do(\Yir\in [\alpha(\ir,l-1),\alpha(\ir,l)])\Big]
        \\
\end{split}
\end{equation} 
for each $1 \leq k \leq L_{\orr}$, $1 \leq l,l' \leq L_{\ir}$, with $l \neq l'$. 
\end{definition}
Notice that the definition of OCE is anti-symmetric in the initial and end level of the intervention variable, meaning that 
\begin{equation*}
    \text{OCE}_{\ir \orr}(k, l \to l')=- \,\text{OCE}_{\ir \orr}(k, l' \to l).
\end{equation*}
In \cref{OCE}, $\Do(\Yir \in [\alpha(\ir,l-1),\alpha(\ir,l)])$ denotes a deterministic intervention on $\Yir$, which consists in setting $\Yir$ to a value $\tilde \yir$ belonging to the interval $[\alpha(\ir,l-1),\alpha(\ir,l)]$. To perform this intervention, we may enforce the %unconditional 
atomic intervention $\Do(\Yir = \tilde \yir)$ with a certain probability distribution $f^*(\tilde \yir)$, called the \emph{intervention policy}, which will modify the distribution of $\Yir$ so that the intervention value $\tilde \yir$ belongs to the interval of interest as following: 
\begin{equation}\label{effectpolicytoy}
\begin{split}
    f(\yor)\Big\rvert_{f^{*}(\tilde \yir)} &= \int_{\mathcal{Y}_\ir} f(\yor \mid \Do(\Yir = \tilde \yir))f^{*}(\tilde \yir)\text{d}\tilde \yir, \\
\end{split} 
\end{equation}
where $\mathcal{Y}_\ir$ represents the support of $Y_\ir$ under intervention, meaning that $f^*$ is any normalized density over the interval $[\alpha(\ir,l-1),\alpha(\ir,l)]$. Given that any distribution of $\Yir$ with support on the previous interval is an eligible intervention policy $f^*(\tilde \yir)$, $\Do(\Yir\in [\alpha(\ir,l-1),\alpha(\ir,l)]$ is not uniquely defined. 

\begin{definition}
    Under the same model of \cref{OCEDef} and given the two intervention policies $f^*(\tilde \yir)$ and $f^{**}( \yir')$ with support respectively on $[\alpha(\ir,l-1),\alpha(\ir,l)]$ and  $[\alpha(\ir,l'-1),\alpha(\ir,l')]$, the ordinal causal effect in \cref{OCEDef} is given by: 
    \begin{equation}\label{OCE2}
\begin{split}
     \text{OCE}_{\ir \orr}(k, l \to l') &= 
     \int_{\alpha(\orr,k-1)}^{\alpha(\orr,k)} \int_{\mathcal{Y}_\ir'} f(\yor \mid \Do(\Yir = \yir'))f^{**}(\yir')\text{d}\yir'\text{d} \yor\\ 
     & \quad - \int_{\alpha(\orr,k-1)}^{\alpha(\orr,k)} \int_{\tilde{\mathcal{Y}}_\ir} f(\yor\mid \Do(\Yir = \tilde \yir))f^{*}(\tilde \yir)\text{d}\tilde \yir\text{d} \yor,
\end{split}
\end{equation} 
for each $1 \leq k \leq L_{\orr}$, $1 \leq l,l' \leq L_{\ir}$, with $l \neq l'$. 
\end{definition}

In \cref{OCE2}, the atomic post-intervention density of $\Yor$ is analytically identifiable through the truncated factorisation formula in  \cref{truncformula}, as stated in the following.   
\begin{proposition}\label{ThePostInt1}
Let $(Y_1, \dots, Y_n) \sim \mathcal{N}(\bm{\mu}, \bm{\Sigma})$ and consider the do operator $\Do(\Yir = \tilde \yir), \, \ir \in \{1, \dots, n\}$. Then the atomic post-intervention distribution of $\Yor,  \, \orr \in \{1, \dots, n\}\setminus \{\ir\}$ is \begin{equation}
    f(\yor \mid \Do(\Yir = \tilde \yir), \bm{\mu},\bm{\Sigma}) = \phi \Big(\mu_{\Do}, \Sigma_{\Do}\Big), 
\end{equation} where
\begin{equation}\label{DOparam}
\begin{split}
    & a=\{\Pa (\ir),\ir\}, \qquad b=\{\orr\},\\
    &\mu_{\Do} =  \mu_{\orr}+(\bm{\Sigma}_{ba}\bm{\Sigma}^{-1}_{aa})_\ir(\yir-\mu_\ir), \\
    &\Sigma_{\text{do}} = \Sigma_{bb} - (\bm{\Sigma}_{ba}\bm{\Sigma}^{-1}_{aa})\bm{\Sigma}_{ab}+(\bm{\Sigma}_{ba}\bm{\Sigma}^{-1}_{aa})_{-\ir} \bm{\Sigma}_{\Pa(\ir) \Pa(\ir)} (\bm{\Sigma}_{ba}\bm{\Sigma}^{-1}_{aa})_{-\ir}^\intercal.
\end{split}
\end{equation}
\end{proposition}
\begin{proof}
    See \cref{ProofTh3}. 
\end{proof}

The previous proposition is equivalent to Prop.~3.1.\ of \citet{castelletti_bayesian_2021}, as we also consider a collection of jointly Gaussian variables described by a DAG. Since the latent Gaussian DAG-model is not miss-specified, for each admissible adjustment set, the atomic post-intervention distribution would result in a normal with parameters $\mu_{\Do}$ and $\Sigma_{\Do}$ depending only on intervention and outcome variables.  The proposition below provides a simplified version of the previous result, incorporating the intervention model.

\begin{proposition}\label{ThePostInt}
Let $(Y_1, \dots, Y_n) \sim \mathcal{N}(\bm{\mu}, \bm{\Sigma})$, with $\bm{\Sigma} = (\bm{I}-\bm{B})^{-1}\bm{V}(\bm{I}-\bm{B})^{-\intercal} $ as in \cref{choldec}. Consider the do operator $\Do(\Yir = \tilde \yir), \, \ir \in \{1, \dots, n\}$. Then the post-intervention distribution of $\Yor ,  \, \orr \in \{1, \dots, n\}\setminus \{\ir\}$ is \begin{equation}\label{disgeninter}
    f(\yor \mid \Do(\Yir = \tilde \yir), \bm{\Sigma}) = \phi\Big(\widetilde{\bm{\mu}}_\orr, \widetilde{\bm{\Sigma}}_{\orr \orr}\Big), 
\end{equation} with 
\begin{equation}
    \begin{split} 
        & \widetilde{\bm{\mu}}_{\orr} =\mu_\orr + \bm{W}_{\orr\ir}\,(\yir-\mu_\ir), \quad 
        \widetilde{\bm{\Sigma}}_{\orr \orr} = \Big[(\bm{I}-\widetilde{\bm{B}})^{-1}\widetilde{\bm{V}}(\bm{I}- \widetilde{\bm{B}})^{-\intercal}\Big]_{\orr\orr},
    \end{split}
\end{equation}
where 
\begin{equation}
    \begin{split}
        & \bm{W}=(\bm{I}-\bm{B})^{-1} \\
        & \widetilde{\bm{V}}=\text{diag}(v_{1}, \dots, v_{\ir-1}, 0, v_{\ir+1}, \dots, v_n) \\
        & \widetilde{\bm{B}}_{hj}= \begin{cases}
            0 \quad &\text{if } \, j=\ir \\
            \bm{B}_{hj} \quad &\text{otherwise} \, 
        \end{cases}%\\
      %  &\widetilde{\bm{W}}=(\bm{I}-\widetilde{\bm{B}})^{-1}. 
    \end{split}
\end{equation}
\end{proposition}
\begin{proof}
    See \cref{ProofTh4}.  
\end{proof}

The result in the proposition is independent of the chosen adjustment set: in general, to compute $\Sigma_{\Do}$ of \cref{DOparam}, it is sufficient to consider the variables preceding the outcome in topological order, and for each of these variables $Y_j $, add the contribution to the variance of the outcome variable along those paths, from $Y_j$ to the outcome, that do not intercept the intervention variable. Proposition \ref{ThePostInt} offers a fast and efficient method for determining the parameters of the post-intervention distribution, and aligns with the interpretation of interventions in the latent space.

Once the atomic post-intervention distribution $f(\yor \mid \Do(\Yir = \tilde \yir))$ is determined through \cref{ThePostInt}, we need to select proper intervention policies $f^*(\tilde \yir)$ and $f^{**}( \yir')$ to compute the ordinal causal effect through \cref{OCE2}. Below we cover two slightly different conceptual strategies.  

One option is to shift the value of the latent variable $\tilde y_\ir$ to any point $y_\ir' \in [\alpha(\ir, l' -1),\alpha(\ir,l')]$, by choosing the same distribution family for both intervention policies, and hereafter we refer to this as the distributional approach. Alternatively, we may define the intervention in the latent space, so that $y_\ir'$ is the corresponding point (in \textit{quantile-sense}) to the point $\tilde y_\ir$, i.e. 
\begin{equation}
    F^{*}\Big[Y_\ir\leq \tilde y_\ir\Big] = F^{**}\Big[Y_\ir \leq y_\ir'\Big], 
\end{equation}
and therefore $y_\ir'$ is a function of $\tilde y_\ir$ 
\begin{equation}\label{changevar}
    y_\ir' = F^{{**}^{-1}}(F^{*}(\tilde y_\ir)).
\end{equation}
Henceforth we refer to the latter strategy as the quantile approach.

The two approaches turn out to be equivalent for the computation of OCE when choosing truncated normals as intervention policies, see \cref{sup-equiv}. The quantile approach extends easily to different choices of intervention policy, though the equivalence with the distributional approach might no longer hold.

The following proposition provides a closed formula to compute OCE in the general case when using a distributional approach and the truncated normal distribution as intervention policy, which seems the most natural choice given that the marginal and the pre-intervention distribution of $\Yir$ is normal.

\begin{proposition}[Computation of the Ordinal Causal Effect]\label{GenDistribution}
 \begin{align}\label{OCEtheorem1}
& \mathrm{OCE}_{\ir \orr}(k, l \to l') \nonumber \\
= &\frac{1}{\Phi(\widebar{\alpha}(\ir,l'))- \Phi(\widebar{\alpha}(\ir,l'-1)}
\quad \begin{aligned}
  \Big[ &\mathcal{BN}\left(\widebar{\alpha}(\ir,l'), \tilde{a}_k, \rho\right)  
  - \mathcal{BN}\left(\widebar{\alpha}(\ir,l'-1), \tilde{a}_k, \rho\right) \\
   & -\mathcal{BN}\left(\widebar{\alpha}(\ir,l'), \tilde{a}_{k-1}, \rho\right) 
   + \mathcal{BN}\left(\widebar{\alpha}(\ir,l'-1), \tilde{a}_{k-1}, \rho\right) \Big]
        \end{aligned}\\
& - \frac{1}{\Phi(\widebar{\alpha}(\ir,l))- \Phi(\widebar{\alpha}(\ir,l-1)} 
  \quad \begin{aligned}
  \Big[ &\mathcal{BN}\left(\widebar{\alpha}(\ir,l), \tilde{a}_k, \rho\right) 
  - \mathcal{BN}\left(\widebar{\alpha}(\ir,l-1), \tilde{a}_k, \rho\right) \\
  &- \mathcal{BN}\left(\widebar{\alpha}(\ir,l), \tilde{a}_{k-1}, \rho\right) 
  + \mathcal{BN}\left(\widebar{\alpha}(\ir,l-1), \tilde{a}_{k-1}, \rho\right)\Big].
        \end{aligned}\label{OCEtheorem2}
        \end{align}
        where $\Phi$ and $\mathcal{BN}$ are respectively the c.d.f.\ of a standard normal distribution and the p.d.f.\ of a standard bivariate normal distribution and 
\begin{equation}
    \begin{split}
        & \widebar{\alpha}(\ir,l) = \frac{\alpha(\ir,l)-\mu_\ir}{\sigma_\ir} \quad \forall \: 1 \leq l \leq L_\ir \quad \text{with} \: \sigma_\ir = \sqrt{\bm{\Sigma}_{\ir \ir}};\\
        &\widebar{\alpha}(\orr,k) = \frac{\alpha(\orr,k)-\widetilde{\bm{\mu}}_\orr}{\sqrt{\widetilde{\bm{\Sigma}}_{\orr \orr}}}=\frac{\alpha(\orr,k)-\mu_\orr}{\sqrt{\widetilde{\bm{\Sigma}}_{\orr \orr}}} - \frac{\bm{W}_{\orr\ir}\sigma_\ir}{\sqrt{\widetilde{\bm{\Sigma}}_{\orr \orr}}}z_\ir = a_k+bz_\ir, \forall 1 \leq k \leq L_\orr, \text{with} \: z_\ir = \frac{\tilde \yir - \mu_\ir}{\sigma_\ir}; \\
        & \tilde{a}_k=\frac{a_k}{\sqrt{1+b^2}}, \quad \rho = -\frac{b}{\sqrt{1+b^2}}. 
    \end{split}
\end{equation}
\end{proposition}

\begin{proof}
    See \cref{ProofTh5}.
\end{proof}

Note that \cref{OCEtheorem1} and \cref{OCEtheorem2} can also be expressed using Owen's T function \citep[formula 3.1]{owenTableNormalIntegrals1980}. Furthermore, for a binary intervention variable, \cref{GenDistribution} can be applied to recover the distributional causal effects of \cref{eqdiseffects}.

In the special case of two binary variables, we can verify, as demonstrated in \cref{AP_Bin}, that our method for defining and computing ordinal causal effects delivers results in agreement with the causal risk differences. However, this equivalence breaks down already for three binary variables due to the impossibility of fully representing their distribution by a latent Gaussian construction, as detailed in \cref{AP_3Var}. 

\section{Experimental Results}
\label{sec4}
In this section, we present simulation studies where we evaluate the performance of our proposed method in estimating ordinal effects for both, known and unknown latent network structures. Finally, we show a real-world application to a psychological dataset \citep{McNally} combining information about obsessive-compulsive disorder and co-morbid depression symptoms. 

For the simulations and analysis, and to implement the computations for evaluating the quantities in the formula of \cref{GenDistribution}, we adopted the \textsf{R} statistical software \citep[v.4.4.1]{statsr}. Details of the different implementation methods and comparisons are provided in \cref{AP_Simulation}.

\subsection{Simulations }
In our simulation study, we generate a random Erd\H{o}s--R\'enyi graph DAG with 16 nodes using the \textsf{randDAG} function from the \textsf{pcalg} package \citep[]{pcalg}, where each node is expected to have 5 neighbours, and illustrated in \cref{AP_Simulation}. We explicitly chose a DAG which is the only element of its Markov equivalence class. The edge weights are sampled uniformly from the intervals \((-1, -0.4)\) and \((0.4, 1)\). Next, we generate the corresponding Gaussian sample $\mathcal{D}_{\bm{Y}}= \{\bm{y}^1, \dots, \bm{y}^N\}$ for the latent variable $\bm{Y}$ with a sample size of \(N = 500\) following the topological order in the DAGs. To ensure identifiability, as already discussed in \cref{seclatentmodel}, we standardized each variable of the Gaussian dataset, centring it at its mean and transforming its covariance matrix into the correlation form.   Lastly, we convert the Gaussian sample into an ordinal sample $\mathcal{D}_{\bm{X}}= \{\bm{x}^1, \dots, \bm{x}^N\}$ of size $N$ for the ordinal variables $\bm{X}$. To this aim, each continuous underlying Gaussian variable $Y_m$ is randomly discretised into ordinal levels through a series of thresholds $\bm{\alpha}_m$, generated from a symmetric Dirichlet distribution \(\text{Dir}(L_m, \nu)\), where \(L_m\) denotes the number of expected ordinal levels for variable \(X_m\), and \(\nu\) is the concentration parameter. In our simulations, \(L_m\) is randomly chosen from the set \([2, 6]\) to get an expected number of levels equal to 4 for each variable, and we set \(\nu = 2\) to avoid levels with very small probabilities. Specifically, the cell probabilities for the ordinal contingency tables of \(X_m\) are first derived using \(\text{Dir}(L_m, \nu)\), and based on these probabilities, we calculate the thresholds for cutting the Gaussian variable in dimension \(m\) using the normal quantile function.  

We regenerate the Gaussian dataset $\mathcal{D}_{\bm{Y}}$ 500 times and cut each of them according to the same original cuts to obtain the corresponding ordinal sample. Furthermore, to gain insights into the variability of plausible causal effects from a given dataset, we implement a non-parametric bootstrap procedure resampling with replacement from the ordinal dataset $\mathcal{D}_{\bm{X}}$ to obtain $M=500$ bootstrapped datasets, while maintaining the sample size $N=500$. For both procedures, we compare the estimates of the causal effects to the theoretical estimates from the known graph and parameters, under two different scenarios, named `Param' and `BN' respectively: 
\begin{enumerate} 
    \item[Param:] refers to a procedure where we assume the graph is known and only estimate the parameters. For the known DAG $\mathcal{G}$, we use OSEM  \citep{luo_learning_2021} to only learn the discretisation thresholds $\bm{\alpha}^j$ and the parameters $\vartheta^j=\cup_{m=1}^n \vartheta_m^j$ for each $j=1, \dots, M$;
    \item[BN:] refers to a procedure where we assume no knowledge of either the graph or the parameters and estimate a complete Bayesian network (`BN'). For unknown graphical structures, we use OSEM  to learn both thresholds $\bm{\alpha}^j=\cup_{m=1}^n \bm{\alpha}^j_m$ and the Bayesian Network $\mathcal{B}^j = (\mathcal{G}^j, \vartheta^j=\cup_{m=1}^n \vartheta_m^j)$ for each $j=1, \dots, M$.
\end{enumerate}

In both cases, we employ the learned thresholds, parameters and either given or learned structure to compute the OCEs.

\begin{figure}[t]
    \centering
    \includegraphics[width=\textwidth]{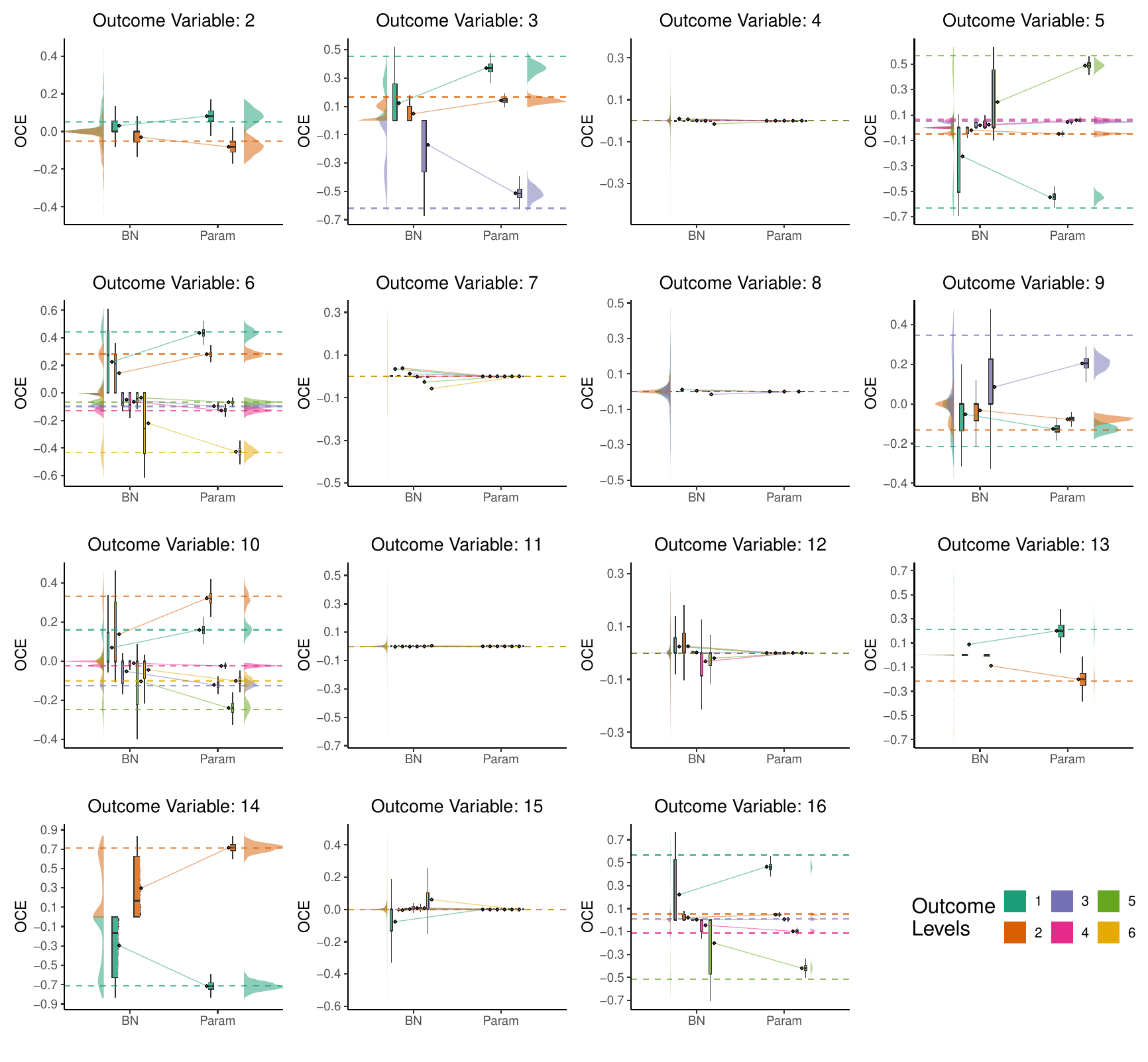}
    \caption{Simulation with Regenerated Data: Ordinal Causal Effects resulting from shifting the intervention variable 1 from its lowest to its upper
level on all other outcome variables. For each level of the outcome, the solid line connects the means of the OCEs (represented by diamond points) across different scenarios, while the dotted line represents the True Causal Effect. }
    \label{matrixfig1R}
\end{figure}

\begin{figure}[t]
    \centering
    \includegraphics[width=\textwidth]{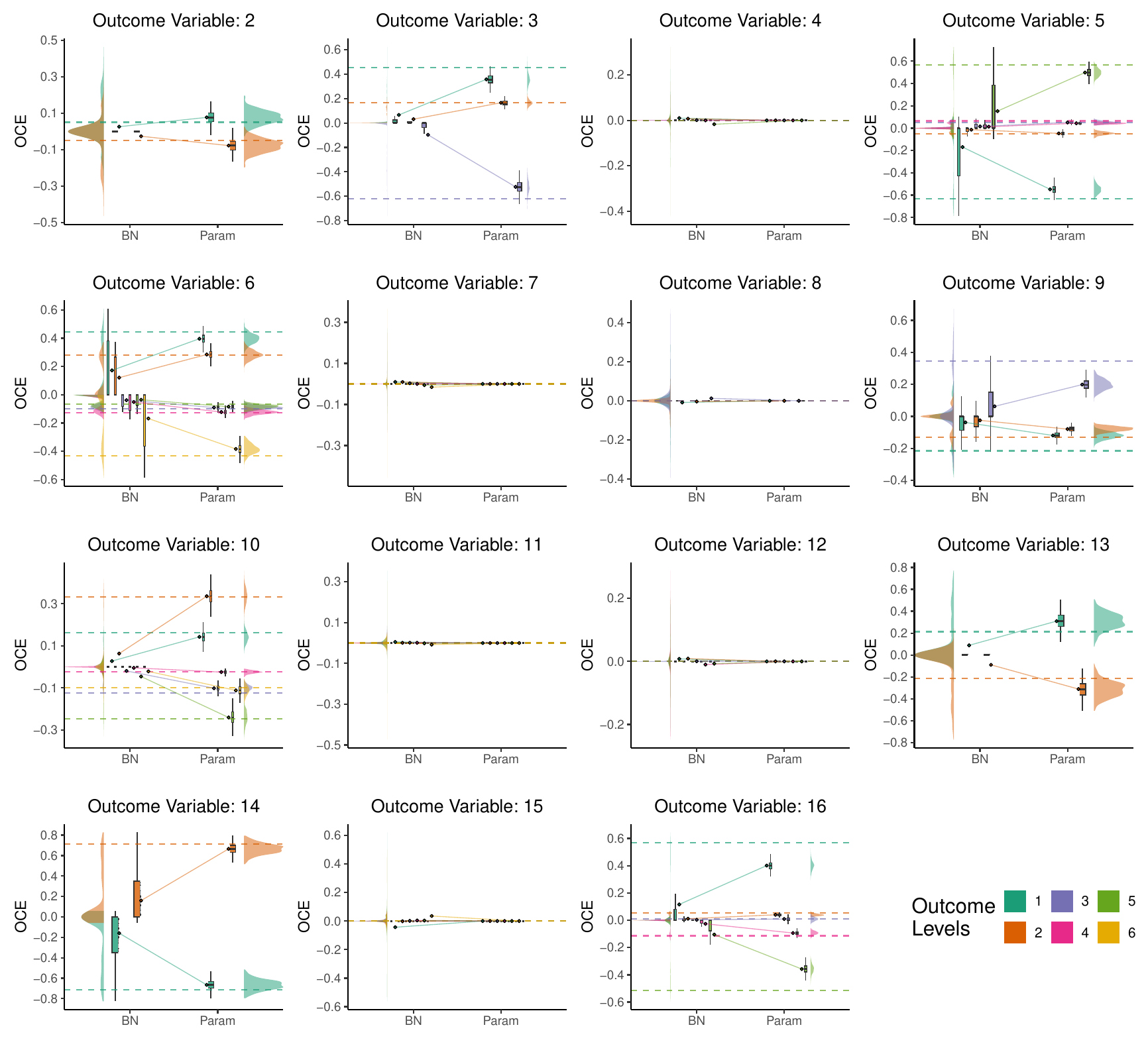}
    \caption{Simulation with Bootstrapped Data: Ordinal Causal Effects resulting from shifting the intervention variable 1 from its lowest to its upper
level on all other outcome variables. For each level of the outcome, the solid line connects the means of the OCEs (represented by diamond points) across different scenarios, while the dotted line represents the True Causal Effect. }
    \label{matrixfig1B}
\end{figure}

To illustrate the performance of our methodology under the two scenarios, we display Raincloud plots \citep{ggrain} of the OCEs resulting from shifting variable 1 from its lowest to its upper level on all the other variables in \cref{matrixfig1R} and in \cref{matrixfig1B} for the simulation with regenerated data and the bootstrapped procedure respectively. For comparison, we also report the theoretical effects determined according to the definition in \cref{GenDistribution} from the parameters used in the simulation to generate the ordinal datasets. Analogous plots for the other root nodes in the True DAG (i.e.\ 8,11,15) are available in \cref{AP_Simulation}. For each pair of intervention and outcome variables among the 16 nodes in the DAG, similar plots are available online (see \cref{sec7}).

As expected, when the graph structure is already known (`Param' approach), our methodology demonstrates improved compatibility in recovering causal effects compared to the `BN' approach, where additional uncertainty about the graph structure is introduced. In both scenarios, the OCEs estimates will depend on the quality of approximation of the underlying covariance matrix, which is obtained through the Monte Carlo EM Algorithm \citep{wei1990} in OSEM \citep{luo_learning_2021}. With increasing sample size, the contingency tables become more reliable, so recovering the original covariance structure through the EM iterations should also gradually improve. 

Overall, the mean causal effects across the two approaches remain largely compatible with the theoretical values, as evidenced by the regenerated data simulation, where the distribution of effects is generally centred on the true effects. Deviations from this pattern, such as in outcome variable 3, can be attributed to deviations in the estimation of the underlying correlation matrix when using the default settings of OSEM. This is particularly the case when these are compounded through paths in the network, as in the case of variable 9, which is a descendant of variable 1 only through variable 3. Improving the accuracy of the correlation matrix estimation could enhance the overall reliability of causal effect estimation within our framework.

\subsection{Analysis of Psychological Data}
In this section, we apply our method to psychological survey data, where preserving the ordinal structure might be particularly valuable for reliably estimating causal effects. We use an ordinal dataset comprising data about 408 adults from \citet{McNally} study on the functional relationships between 10 five-level symptoms of obsessive-compulsive disorder (OCD) and 16 four-level depression symptoms \citep{McNally}, measured with the self-report Yale-Brown Obsessive-Compulsive Scale (Y-BOCS-SR) \citep{Steketee} and the Quick Inventory of Depressive Symptomatology (QIDS-SR) \citep{Rushetal} respectively. Because two pairs of variables in the original depression dataset fundamentally encode the same information, we followed the approach in \citet{luo_learning_2021} to combine each pair in a single variable with seven levels. Details of the symptoms included in the latent DAG-model are summarised in Supplementary Table S2 of \citet{luo_learning_2021}. 

We followed the procedure in \citet{luo_learning_2021} to obtain DAG estimates from 500 bootstrap samples of the data by running the OSEM algorithm with Monte Carlo sample size $K=5$ and penalty coefficient $\lambda = 6$. 

\begin{figure}[t]
    \centering
\includegraphics[width=\textwidth]{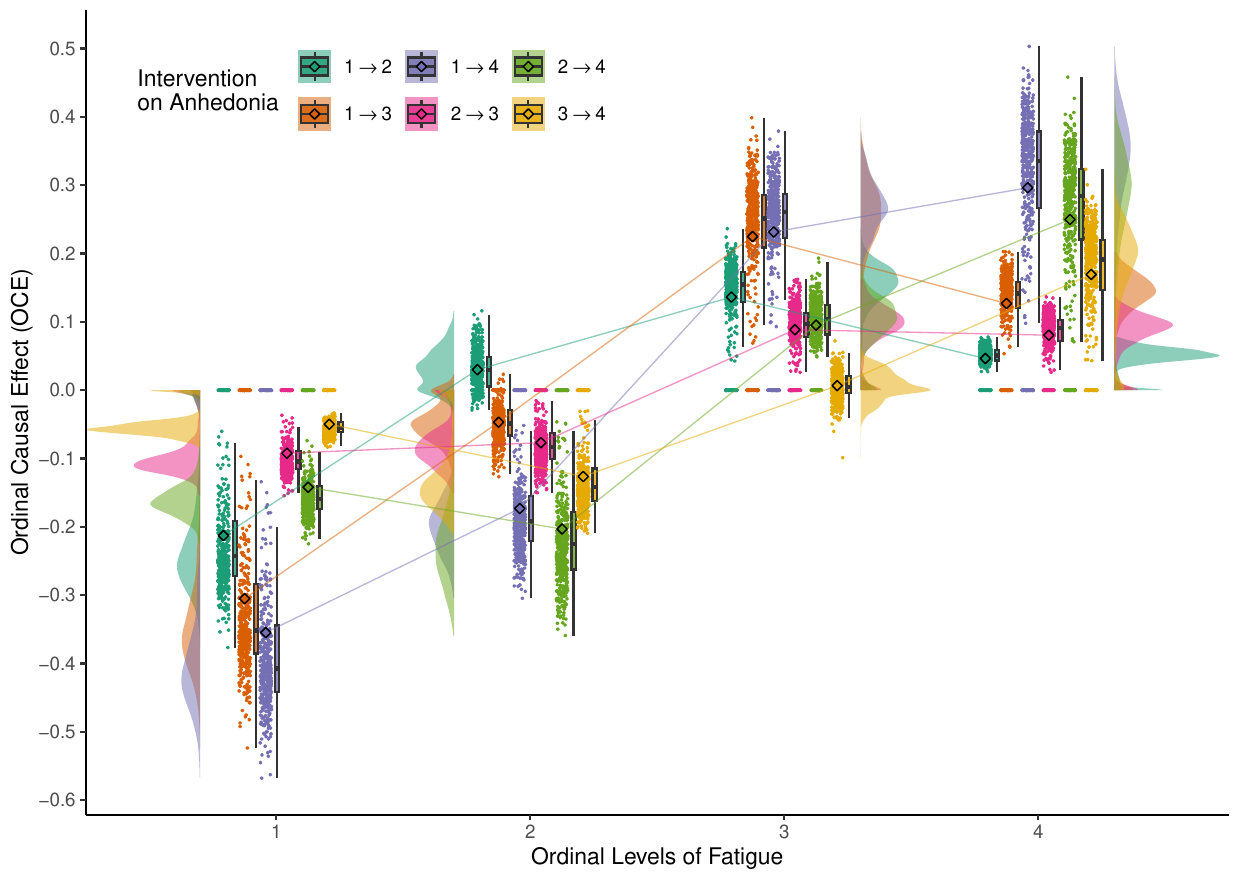}
    \caption{Ordinal Causal Effect of Anhedonia (variable 9) on Fatigue (variable 10). The solid line connects the means of the OCEs (represented by diamond points) across different levels of the outcome variable, for each possible shift of the intervention variable.}
    \label{Effect910}
\end{figure}

As a visual representation of the bootstrapped estimates, we display the adjacency matrices of the DAGs obtained via OSEM converted to CPDAGs in a heatmap in \cref{Heatmap} of \cref{AP_Simulation}, where the intensity of each cell is proportional to how often each edge appears in the bootstrapped sample. Further, we investigate the causal relationship along the direct edge most frequently appearing in the 500 bootstrapped CPDAGs, by deriving the ordinal causal direct effects of anhedonia (variable 9) on fatigue (variable 10) in the DAGs in the sample. Raincloud plots including histograms and boxplots of the estimated effects appear in \cref{Effect910}. Finally, we also display raincloud plots summarising the effects of shifting variable 9 from its lowest to its upper level on all other variables and for all their levels in \cref{Effect9} of Appendix. Analogous plots for all other possible choices of intervention variables are available online (see \cref{sec7}). 

\section{Discussion}
\label{sec5}
In this work, we address the challenge of defining and estimating causal effects from ordinal data within a latent Gaussian DAG framework. Modelling each ordinal variable as a marginal discretisation of an underlying Gaussian variable has the appeal of preserving the ordinality among categories. Additionally, the construction via a latent continuous multivariate Gaussian distribution enables us to propose a definition of Ordinal Causal Effects starting from pointwise intervention distributions in the latent space. Given the latent variable construction, we could derive closed-form expressions for the OCEs. In the special case of two binary variables, our approach retrieves the causal effects corresponding to the traditional definition.

By evaluating the shift in probability for all ordinal levels, our framework can recover any estimand of interest, comparing the probabilities that an outcome variable takes different levels when intervening to shift an intervention variable from any ordinal level to any other. In particular, the framework incorporates the causal estimand presented by \citet{lupeng} and defined as the shift in the probability of falling above a certain category between the two levels of a binary intervention variable. This estimand can be readily derived by simply adjusting the outer limits of integration in \cref{OCE2}.

Given the marginal normality of latent variables, we adopt the most natural choice of intervention policy, selecting a truncated normal distribution supported by the band representing the level of the ordinal variable intended as the target of the intervention. The truncated normal reflects the latent Gaussian model, but other intervention policies would be possible. For example, one could also use bounded distributions, such as the uniform. However, while bounded policies work well for inner intervention levels, they struggle to adequately describe interventions targeting semi-infinite intervals like $(-\infty, a)$ or $(b, +\infty)$ with $a, b \in \R$. The truncated normal policy on the other hand remains well-suited even for intervention targeting the lowest and highest levels of an ordinal intervention variable. 

In our simulation studies, we estimate the parameters of a Bayesian network, starting either from a known or an estimated graph structure, using the OSEM algorithm. From the estimated parameters we could compute the OCEs and compare them with those derived from the true underlying parameters, observing good agreement. Furthermore, we demonstrate the applicability of our methodology by analysing psychological survey data, where we estimate a network and compute OCEs while illustrating their variability using bootstrapping.

Learned DAGs may belong to various Markov equivalence classes, while different DAGs within the same equivalence class may yield distinct causal effects due to variations in the parent sets of the intervened upon node. To account for the various possibilities within an equivalence class, \citet{maathuis_estimating_2009} proposed to find bounds on the causal effects by examining all possible effects implied by the DAGs in the class. However, since exhaustively enumerating all DAGs within a CPDAG can be computationally prohibitive, even for small networks, they suggested a strategy to focus on the distinct causal effects within a CPDAG. Alternatively, a Bayesian approach can be employed to sample Bayesian networks and estimate the posterior distribution of causal effects \citep{moffa2017psychoBayesian, kuipers2022efficient, Viinikka2020}. This approach automatically accounts for the general uncertainty in the DAG structure, and it yields effect distributions over the Markov equivalence classes assuming that all structures in a class have the same probability. For ordinal variables, \citet{grzegorczyk_being_2024} recently proposed a method for sampling networks. Subsequently, one can estimate OCEs for each sampled DAG using our proposed methodology and obtain the posterior distribution of causal effects. 

While we focus on single interventions, it is important to note that in real situations any exogenous intervention may simultaneously affect multiple target variables. Consequently, we may wish to predict how joint interventions will affect an outcome variable. Since the methods in  \citet{Viinikka2020} extend easily to deal with multiple interventions, then under the latent Gaussian model, accounting for joint interventions, akin to \citet{nandyEstimatingEffectJoint2017}, will be a natural extension to the latent space. Our approach would then provide the framework to map this to multiple ordinal interventions through an extension of the post-intervention distribution illustrated in \cref{ThePostInt}.

An alternative strategy consists of estimating intervention effects directly on the underlying continuous constructs since there are real-world situations where such constructs appear more natural, as may be the case in measures of quality of life. However, the non-identifiability of the underlying Gaussian distribution, without additional constraints, complicates interpretability compared to defining effects directly linked to the ordinal scale.

Finally, our methodology could extend to estimating causal effects in mixed data settings, where both continuous and ordinal variables are present. This setting can be modelled as a DAG-based Gaussian structure, with some variables discretised and others remaining continuous. While dealing with mixed continuous and nominal categorical variables would require further extension of the OSEM algorithm for estimating graphical structures and parameters, our framework could form the basis for mapping from the latent to the observed space, and hence for estimating causal effects.

\section{Software and Supplementary Materials}
\label{sec7}

Software in the form of \textsf{R} code, together with additional simulation results and complete documentation is available at
\url{https://github.com/martinascauda/OrdinalEffects}.

\acks{The authors thank Dr.\ Enrico Giudice and Dr.\ Ching Wong for helpful discussions and the University of Basel for hosting Martina Scauda during her Master's Thesis, funded by 
the Swiss-European Mobility Program. 
}


\vskip 0.2in
\bibliography{refs}

\newpage
\appendix
\section{Additional Theoretical Results}
\subsection{A Toy Model}\label{apptoymodel}
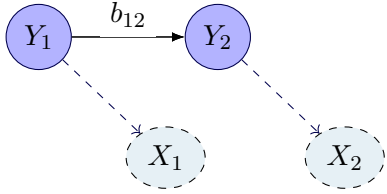
\begin{figure}[h!]
    \centering
    \begin{tikzpicture}[scale=0.2,baseline=(current bounding box.north), state/.style={circle, draw=blue!30!black, fill=blue!30, minimum size=2em}]
    \node[state] (m) at (0,0) {$Y_1$};
    \node[state] (h) [right =of m, shift={(.5,0)}] {$Y_2$};
    \node[da, fill=blue!60!green!10] (k) [below right =of m] {$X_1$};
    \node[da, fill=blue!60!green!10] (b) [below right = of h] {$X_2$};
    \path (m) edge (h);
    \draw (m) -- (h) node [midway, fill=white] {$b_{12}$};
    \draw[dashed, draw=blue!30!black, ->] (m) -- (k);
    \draw[dashed, draw=blue!30!black, ->] (h) -- (b);
  \end{tikzpicture} 
  % \begin{tikzpicture}
  %   \node[state] (m) at (0,0) {$Y_1$};
  %   \node[state] (j) [right =of m] {$Y_2$};
  %   \node[da] (k) [below =of m] {$X_1$};
  %   \node[da] (v) [below =of j] {$X_2$};
  %   \draw[edge] (m) to node[left, sloped,xshift=8pt,yshift=4pt]{\footnotesize 
  %   $b_{12}$}(j);
  %   \draw[dashed, ->] (m) -- (k);
  %   \draw[dashed, ->] (j) -- (v);
  % \end{tikzpicture} 
  \caption{Toy model on DAG $\mathcal{G}$. }
  \label{toymodel}
\end{figure}

For illustrative purposes, we first provide a definition of causal effect of an intervention in a toy Gaussian DAG-model, where there are just two binary variables $X_1$ and $X_2$, assumed to be obtained by marginally discretising two underlying Gaussian variables $Y_1$ and $Y_2$. Let the parameters $\theta$ and $\bm{\alpha}$ of the model be given together with the DAG $\mathcal{G}$, illustrated in \cref{toymodel}.   

A measure of the causal effect on $X_2$ of an intervention on $X_1$ may be the following  
\begin{equation}\label{goal1}
       \Prob[X_2 = \tau(2,k) \mid \Do(X_1 = \tau(1,l'))] -
    \Prob[X_2 = \tau(2,k) \mid \Do(X_1 = \tau(1,l))] 
\end{equation}
for each $l \neq l'$ and $l,l',k \in \{1,2\}$. Following \cref{OCEDef}, \cref{goal1} can be equivalently expressed on the latent scale as 
\begin{equation}\label{toymodeleffect}
\begin{split}
     \Prob\Big[Y_2 &\in [\alpha(2,k-1),\alpha(2,k)]\mid \Do(Y_1 \in [\alpha(1,l'-1),\alpha(1,l')])\Big]\\
        &\quad -  \Prob\Big[Y_2 \in [\alpha(2,k-1),\alpha(2,k)] \mid \Do(Y_1 \in [\alpha(1,l-1),\alpha(1,l)])\Big]
\end{split}
\end{equation}
A closed formula to compute it, using either distributional or quantile approach, is provided in \cref{sup-equiv}. 
\subsection{Proof of  \texorpdfstring{\cref{ThePostInt1}}{ThePostInt1}}\label{ProofTh3}
\begin{proof}
Let $$\bm{Y}_a =  \left[ \begin{array}{c}
         \bm{\Pa}(\ir)\\ \hline 
         \Yir
    \end{array} \right] \quad \text{and} \quad Y_b = \begin{bmatrix} \Yor \end{bmatrix}$$ be marginally distributed as $ (\bm{Y}_a, Y_b) \sim \mathcal{N}(\widebar{\bm{\mu}}, \widebar{\bm{\Sigma}})$ with block-wise mean and covariance defined as $$\widebar{\bm{\mu}} = \left[ \begin{array}{c}
         \bm{\mu}_a \\\hline
         \mu_b
    \end{array}
    \right] \quad \widebar{\bm{\Sigma}} = \left[\begin{array}{c|c}
         \bm{\Sigma}_{aa}& \bm{\Sigma}_{ab}  \\
         \hline
          \bm{\Sigma}_{ba}& \Sigma_{bb}
    \end{array} \right]$$ and $|\bm{\Sigma}_{aa}| > 0.$
    It follows from classic results on conditional normal distribution that $Y_b \mid \bm{Y_a}= \bm{y}_a\sim \mathcal{N}(\mu_{b\mid a}, \Sigma_{b \mid a})$ is a univariate distribution with 
    \[\mu_{b\mid a} = \mu_b + \bm{\Sigma}_{ba}\bm{\Sigma}^{-1}_{aa}(\bm{y}_a - \bm{\mu}_a)\]
    \[\Sigma_{b \mid a} = \Sigma_{bb} - (\bm{\Sigma}_{ba}\bm{\Sigma}^{-1}_{aa})\bm{\Sigma}_{ab}.  \]
    Here $\bm{\Sigma}^{-1}_{aa}$ is the generalised inverse of $\bm{\Sigma}_{aa}$ and $\bm{\Sigma}_{b \mid a}$ is the Shur complement of $\bm{\Sigma}_{aa}$ in $\widebar{\bm{\Sigma}}$.

Call
\[\hat \yor=\yor-\mu_b-(\bm{\Sigma}_{ba}\bm{\Sigma}^{-1}_{aa})_\ir(\yir-\mu_\ir)\]
and denoting $\bm{\Pa}(\ir)$ with $\p$, 
\[Z =(\bm{\Sigma}_{ba}\bm{\Sigma}^{-1}_{aa})_{-i} \quad \quad C = Z^\intercal \Sigma^{-1}_{b \mid a}Z+\bm{\Sigma}^{-1}_{pp} \quad \quad u^\intercal=\Sigma^{-1}_{b \mid a}\hat \yor^\intercal Z \quad \tilde{\p}=\p - \bm{\mu}_p \]
where $(\bm{\Sigma}_{ba}\bm{\Sigma}^{-1}_{aa}) = \left[ \begin{array}{c}
         (\bm{\Sigma}_{ba}\bm{\Sigma}^{-1}_{aa})_{-i} \\\hline
         (\bm{\Sigma}_{ba}\bm{\Sigma}^{-1}_{aa})_{i} 
    \end{array}
    \right] $ is a $(p+1)\times 1$ vector with $p$ the number of parent nodes of the intervention variable.  

It holds
\begin{equation}\label{intervention}
    \begin{split}
        &\int_{\R^p} f(\yor \mid \yir, \p)\, \tilde{f}(\p) \text{d}\p \\ & \, =\frac{1}{\sqrt{2\pi \Sigma_{b \mid a}}}\cdot\frac{1}{(2\pi)^{\frac{d}{2}}|\bm{\Sigma}_{pp}|^{\frac{1}{2}}}\int_{\R^p} \exp\Big\{-\frac{1}{2}(\yor-\bm{\mu}_{b\mid a})^\intercal \Sigma_{b \mid a}^{-1}(\yor-\bm{\mu}_{b\mid a})-\frac{1}{2}(p-\bm{\mu}_p)^\intercal\bm{\Sigma}_{pp}^{-1}(p-\bm{\mu}_p)\Big\}\text{d}\p \\
        & \propto \int_{\R^{p}} \exp\Big\{-\frac{1}{2} \Big(\hat \yor -(\bm{\Sigma}_{ba}\bm{\Sigma}^{-1}_{aa})_{-i}(p-\bm{\mu}_p)\Big)^\intercal \Sigma_{b \mid a}^{-1}\Big(\hat \yor -(\bm{\Sigma}_{ba}\bm{\Sigma}^{-1}_{aa})_{-i}(p-\bm{\mu}_p)\Big) \\& - \frac{1}{2}(p-\bm{\mu}_p)^\intercal\bm{\Sigma}_{pp}^{-1}(p-\bm{\mu}_p)\Big\}\text{d}\p \\
        & \,\propto \exp\Big\{-\frac{1}{2}\hat \yor^\intercal \Sigma_{b \mid a}^{-1}\hat \yor \Big\}\int_{\R^p} \exp \Bigg\{-\frac{1}{2}\Big[(p-\bm{\mu}_p)^\intercal C (p-\bm{\mu}_p) -2\Sigma_{b \mid a}^{-1}\hat \yor^\intercal Z(p-\bm{\mu}_p) \Big]\Bigg\}\text{d}\p \\
        & \, \underset{(*)}{\propto} \exp\Big\{-\frac{1}{2}\hat \yor^\intercal \Sigma_{b \mid a}^{-1}\hat \yor \Big\}\int_{\R^p} \exp\Bigg\{-\frac{1}{2}\Big[(\tilde \p -C^{-1}u)^\intercal C (\tilde \p -C^{-1}u) -u^\intercal C^{-1}u\Big]\Bigg\}\text{d}\tilde \p \\
        & \, \propto \exp\Big\{-\frac{1}{2}\Big(\hat \yor^\intercal \Sigma_{b \mid a}^{-1}\hat \yor -u^\intercal C^{-1}u\Big)\Big\} \\
        & \, \propto \exp\Big\{-\frac{1}{2}\hat \yor^\intercal \Big(\Sigma^{-1}_{b \mid a} -(\Sigma^{-1}_{b \mid a})ZC^{-1}Z^\intercal (\Sigma^{-1}_{b \mid a})\Big)\hat \yor\Big\}
    \end{split}
\end{equation}
Notice that the domain of the integral is  translational invariance, therefore in (*) one could consider $\tilde{\p}$ instead of $\p$. In addition, since the matrix $C$ is positive definite, the integral $$\int_{\R^p} \exp\Bigg\{-\frac{1}{2}\Big[(\tilde \p -C^{-1}u)^\intercal C (\tilde \p -C^{-1}u)\Big]\Bigg\}\text{d}\tilde \p \propto 1 $$ because it is the kernel of a Gaussian distribution with mean $C^{-1}u$ and covariance matrix $C^{-1}$  integrated over its support. 

From \cref{intervention}, it follows that the kernel of the distribution of $\hat \yor \mid \Do(\Yir = \yir)$ is 
\[\exp\Big\{-\frac{1}{2}\hat \yor^\intercal \Big(\Sigma^{-1}_{b \mid a} -(\Sigma^{-1}_{b \mid a})ZC^{-1}Z^\intercal (\Sigma^{-1}_{b \mid a})\Big)\hat \yor\Big\}\]
which corresponds to a $\mathcal{N}(0,\Sigma_{\text{do}})$ with $$\Sigma_{\text{do}}^{-1}=\Sigma^{-1}_{b \mid a} -(\Sigma^{-1}_{b \mid a})ZC^{-1}Z^\intercal (\Sigma^{-1}_{b \mid a}).$$

The expression of $\Sigma_{\text{do}}$ can be extracted from the previous equation using the Sherman-Morrison-Woodbury formula \citep[][page 258]{woodbury}, which states that the inverse of a rank-k modification of a matrix $A$ can be computed by performing a rank-k correction to the inverse $A^{-1}$. Precisely, the Woodbury formula is  
\begin{equation}\label{wood}
    A^{-1}-A^{-1}U(B^{-1}+VA^{-1}U)^{-1}VA^{-1} = (A+UBV)^{-1}.
\end{equation}
where $A,U,C$ and $V$ are conformable matrices: $A$ is $d \times d$, $B$ is $k \times k$, $U$ is $d \times k$ and $V$ is $k \times d$.

Take $$ A=\Sigma_{b \mid a}, \quad U=Z,\quad B=\bm{\Sigma}_{pp}, \quad V=Z^\intercal,$$ it follows 
\begin{equation}
\begin{split}
\Sigma_{\text{do}}^{-1} &=
    \Sigma^{-1}_{b \mid a} -(\Sigma^{-1}_{b \mid a})Z\Big(\bm{\Sigma}^{-1}_{pp}+Z^\intercal \Sigma^{-1}_{b \mid a}Z\Big)^{-1}Z^\intercal (\Sigma^{-1}_{b \mid a}) \\ 
    &=(\Sigma_{b \mid a}+Z\bm{\Sigma}_{pp}Z^\intercal)^{-1}= (\Sigma_{b \mid a}+(\bm{\Sigma}_{ba}\bm{\Sigma}^{-1}_{aa})_{-i} \bm{\Sigma}_{pp} (\bm{\Sigma}_{ba}\bm{\Sigma}^{-1}_{aa})_{-i}^\intercal)^{-1}.
\end{split}
\end{equation}
Recalling $\hat \yor=\yor-\mu_b-(\bm{\Sigma}_{ba}\bm{\Sigma}^{-1}_{aa})_\ir(\yir-\mu_\ir)$, the atomical post-intervention distribution is
\begin{equation}
    \yor \mid \Do(\Yir = \yir) \sim \mathcal{N}(\mu_{\text{do}},\Sigma_{\text{do}})
\end{equation}
with 
\begin{equation}
\begin{split}
    &\mu_{\text{do}} =  \mu_\orr+(\bm{\Sigma}_{ba}\bm{\Sigma}^{-1}_{aa})_\ir(\yir-\mu_\ir) \\
    &\Sigma_{\text{do}} = \Sigma_{bb} - (\bm{\Sigma}_{ba}\bm{\Sigma}^{-1}_{aa})\bm{\Sigma}_{ab}+(\bm{\Sigma}_{ba}\bm{\Sigma}^{-1}_{aa})_{-i} \bm{\Sigma}_{pp} (\bm{\Sigma}_{ba}\bm{\Sigma}^{-1}_{aa})_{-i}^\intercal.
\end{split}
\end{equation}

% Could be nice: show that the derived atomical post-intervention distribution is equivalent to the results obtained from \citep{castelletti_bayesian_2021}.
\end{proof}

\subsection{Proof of \texorpdfstring{\cref{ThePostInt}}{ThePostInt}}\label{ProofTh4}
Remember that the marginal variance-covariance matrix $\widebar{\bm{\Sigma}}$ of $\widebar{\bm{Y}}=(\bm{pa}(\ir), \yir,\yor)$ is obtained selecting from $\bm{\Sigma}$ the appropriate subset corresponding to the variables $\widebar{\bm{Y}}$. Since $\widebar{\bm{\Sigma}}$ is still symmetric and positive definite, it also admits a Cholesky factorisation. This decomposition of $\widebar{\bm{\Sigma}}$ can be obtained as rank $n-(p+2)$ downdate of the Cholesky decomposition of $\bm{\Sigma}$, because $n-(p+2)$ rows and columns from $\bm{\Sigma}$ are deleted. However, this procedure is rather complex to perform as the rank of the downdate grows. Therefore, an alternative approach, involving linear SEM properties, is involved in the following proof. 

\begin{proof}
Without loss of generality,  assuming topological ordering, let  $$\widebar{\bm{\Sigma}} = \widebar{\bm{W}}\, \widebar{\bm{V}}^{1/2} (\widebar{\bm{W}}\, \widebar{\bm{V}}^{1/2} )^\intercal$$ be the marginal Cholesky factorisation of $\widebar{\bm{\Sigma}}$, the marginal variance-covariance matrix of $\widebar{\bm{Y}}=(\bm{\Pa}(\ir), \yir,\yor)$.

Recall that from \citet{wright}, $$(\bm{W})_{jh} = (\bm{I}-\bm{B})_{jh}$$ is the total causal effect of $h$ on $j$ and the diagonal of $\bm{W}$ is made by 1's. Integrating out variables not in $\widebar{\bm{Y}}$ increases the variance, i.e. $\widebar{\bm{V}}_{jj} \geq \bm{V}_{jj}$ for all $j \in \widebar{\bm{Y}}$, but it does not affect the total effects between remaining variables $\widebar{\bm{Y}}$. Hence,  $\widebar{\bm{W}}$ is just the sub-matrix $\bm{W}_{\widebar{\bm{Y}}, \widebar{\bm{Y}}}$of $\bm{W}$ corresponding to variables in $\widebar{\bm{Y}}$, i.e.
\begin{equation}\label{wbar}
    \widebar{\bm{W}}_{hj} = \bm{W}_{hj}\quad \text{for all } h,j \in \{\Pa(\ir),\ir,\orr\}.
\end{equation}

From \cref{choldec}, the elements of the variance in the marginal Cholesky decomposition are given by $$\widebar{\bm{V}}_{\ir \ir}=0$$ (in fact, $\Yir$ is enforced to assume the constant value $\yir$ by the do operator) and for $j \in \{\Pa(\ir),\orr\}$ by 
\begin{equation}
    \widebar{\bm{V}}_{jj} = \bm{V}_{jj} + \sum_{k < j} \widehat{\bm{W}}_{jk}^2 \bm{V}_{kk}
\end{equation}
where $Y_k$ is one of the marginalized nodes and \begin{equation}\label{pathswithout}
\widehat{\bm{W}}_{jk} = \bm{W}_{jk}- \sum_{k<h<j} \bm{W}_{hk}\bm{W}_{jh}
\end{equation}
with $Y_h \in \widebar{\bm{Y}}$. Here, $\widehat{\bm{W}}_{jk}$ is the weighted (by the path coefficients) number of paths that goes from $Y_k$ to $Y_j$ and do not pass through one of the already considered nodes $Y_h \in \widebar{\bm{Y}}$ (since the increase in variance carried by that path has already been considered in $\widebar{V}_{hh}$).

Recall $\bm{Y}_a=(\bm{Y}_{\bm{pa}(\ir)}, \Yir)$ and $\bm{Y}_b=\Yor$, writing $\widebar{\bm{W}}$ and $\widebar{\bm{V}}$ as block matrix 
\[ \widebar{\bm{W}} = \begin{bmatrix}
    \bm{W}_{aa} & \bm{0} \\
    \bm{W}_{ba} & W_{bb}
\end{bmatrix} \quad \widebar{\bm{V}} = \begin{bmatrix}
    \widebar{\bm{V}}_{aa} & \bm{0}\\
    \bm{0} & \widebar{V}_{bb}
\end{bmatrix},\]
it follows
\begin{equation}
    \widebar{\bm{\Sigma}} = \begin{bmatrix}
    \bm{\Sigma}_{aa} &  \bm{\Sigma}_{ab} \\
    \bm{\Sigma}_{ba} &  \bm{\Sigma}_{bb}
\end{bmatrix} = \begin{bmatrix}
\bm{W}_{aa}\widebar{\bm{V}}_{aa}\bm{W}_{aa}^\intercal & \bm{W}_{aa}\widebar{\bm{V}}_{aa}\bm{W}_{ba}^\intercal \\
\bm{W}_{ba}\widebar{\bm{V}}_{aa}\bm{W}_{aa}^\intercal & \bm{W}_{ba}\widebar{\bm{V}}_{aa}\bm{W}_{ba}^\intercal+W_{bb}\widebar{V}_{bb}W_{bb}^\intercal
\end{bmatrix}. 
\end{equation}

Consider the parameters in \cref{DOparam} of the atomical post-intervention distribution
\begin{equation}
\begin{split}
    &\widetilde{\mu}_{o} =  \mu_\orr+(\bm{\Sigma}_{ba}\bm{\Sigma}^{-1}_{aa})_\ir(\yir-\mu_\ir) \\
    &\widetilde{\bm{\Sigma}}_{\orr \orr} = \Sigma_{bb} - (\bm{\Sigma}_{ba}\bm{\Sigma}^{-1}_{aa})\bm{\Sigma}_{ab}+(\bm{\Sigma}_{ba}\bm{\Sigma}^{-1}_{aa})_{-i} \bm{\Sigma}_{pp} (\bm{\Sigma}_{ba}\bm{\Sigma}^{-1}_{aa})_{-i}^\intercal.
\end{split}
\end{equation}
Then 
\[(\bm{\Sigma}_{ba}\bm{\Sigma}^{-1}_{aa}) = \bm{W}_{ba}\widebar{\bm{V}}_{aa}\bm{W}_{aa}^\intercal \bm{W}_{aa}^{-\intercal}\widebar{\bm{V}}^{-1}_{aa}\bm{W}_{aa}^{-1} = \bm{W}_{ba}\bm{W}_{aa}^{-1}. \]
Since $W_{bb}=1$, this last quantity corresponds to the last row, except last element, of the inverse $\widebar{\bm{W}}^{-1}$, which is trivial to compute given the block matrix form. In fact,
\[\widebar{\bm{W}} = \begin{bmatrix}
    \bm{W}_{aa} & \bm{0} \\
    \bm{W}_{ba} & W_{bb}
\end{bmatrix}= \begin{bmatrix}
    \bm{I} & \bm{0} \\
    \bm{0} & W_{bb}
\end{bmatrix}\begin{bmatrix}
    \bm{I} & \bm{0} \\
    W_{bb}^{-1}\bm{W}_{ba}\bm{W}_{aa}^{-1} & \bm{I}
\end{bmatrix} \begin{bmatrix}
    \bm{W}_{aa} & \bm{0} \\
    \bm{0} & I
\end{bmatrix},  \]
hence
\[\widebar{\bm{W}}^{-1} = \begin{bmatrix}
    \bm{W}_{aa}^{-1} & \bm{0} \\
    -W_{bb}^{-1}\bm{W}_{ba}\bm{W}_{aa}^{-1} & W_{bb}^{-1}
\end{bmatrix}= \begin{bmatrix}
    \bm{W}_{aa}^{-1} & \bm{0} \\
    \bm{0} & I
\end{bmatrix}\begin{bmatrix}
    \bm{I} & \bm{0} \\
    -W_{bb}^{-1}\bm{W}_{ba}\bm{W}_{aa}^{-1} & \bm{I}
\end{bmatrix}\begin{bmatrix}
    \bm{I} & \bm{0} \\
    \bm{0} & W_{bb}^{-1}
\end{bmatrix} \]

In particular, exploiting again the block form of matrixes, it holds
\[\bm{W}_{ba}= \begin{bmatrix}
    \bm{W}_{op} & W_{oi}
\end{bmatrix} \quad \bm{W}_{bb} = \begin{bmatrix}
    \bm{W}_{pp} & 0 \\
    \bm{W}_{ip} & W_{\ir \ir}
\end{bmatrix} \quad \bm{W}_{bb}^{-1} = \begin{bmatrix}
    \bm{W}_{pp}^{-1} & 0 \\- \bm{W}_{ip}\bm{W}_{pp}^{-1} & w_{\ir \ir}^{-1}
\end{bmatrix}\]
and therefore 
\[(\bm{\Sigma}_{ba}\bm{\Sigma}^{-1}_{aa}) = \begin{bmatrix}
    (\bm{\Sigma}_{ba}\bm{\Sigma}^{-1}_{aa})_{-i} & (\bm{\Sigma}_{ba}\bm{\Sigma}^{-1}_{aa})_\ir
\end{bmatrix}= \bm{W}_{ba}\bm{W}_{aa}^{-1} = \begin{bmatrix}
    \bm{W}_{op}\bm{W}_{pp}^{-1}-W_{oi}\bm{W}_{ip} \bm{W}_{pp}^{-1} & W_{oi} 
\end{bmatrix}.\]
This means that $$\widetilde{\mu}_{o} =  \mu_\orr+(\bm{\Sigma}_{ba}\bm{\Sigma}^{-1}_{aa})_\ir(\yir-\mu_\ir)= \mu_\orr+W_{oi}(\yir-\mu_\ir).$$

Regarding $\widetilde{\bm{\Sigma}}_{\orr \orr}=\Sigma_{bb} - (\bm{\Sigma}_{ba}\bm{\Sigma}^{-1}_{aa})\bm{\Sigma}_{ab}+(\bm{\Sigma}_{ba}\bm{\Sigma}^{-1}_{aa})_{-i} \bm{\Sigma}_{pp} (\bm{\Sigma}_{ba}\bm{\Sigma}^{-1}_{aa})_{-i}^\intercal$, since $W_{bb}=1$, notice that the terms 
\begin{equation*}
    \begin{split}
    &\Sigma_{bb}-(\bm{\Sigma}_{ba}\bm{\Sigma}^{-1}_{aa})\bm{\Sigma}_{ab} = \bm{W}_{ba}\widebar{\bm{V}}_{aa}\bm{W}_{ba}^\intercal+\widebar{V}_{bb}- \bm{W}_{ba}\bm{W}_{aa}^{-1}\bm{W}_{aa}\widebar{\bm{V}}_{aa}\bm{W}_{ba}^\intercal  = \widebar{V}_{bb} \\
    &(\bm{\Sigma}_{ba}\bm{\Sigma}^{-1}_{aa})_{-i} \bm{\Sigma}_{pp} (\bm{\Sigma}_{ba}\bm{\Sigma}^{-1}_{aa})_{-i}^\intercal = (W_{op}-W_{oi}\bm{W}_{ip})\widebar{\bm{V}}_{pp}(W_{op}-W_{oi}\bm{W}_{ip})^\intercal= \widetilde{ \bm{W}}_{op}\widebar{\bm{V}}_{pp}\widetilde{ \bm{W}}_{op}^\intercal.
    \end{split}
\end{equation*}
where $\widetilde{\bm{W}}_{\orr p}$ is a vector $ p \times 1$, representing the weighted (by the path coefficients) number of paths that goes from $Y_p$ to $\Yor$ and do not pass through the intervention variable $\Yir$. 

In conclusion \begin{equation}
    \widetilde{\bm{\Sigma}}_{\orr \orr} = \widebar{V}_{\orr \orr} + \widetilde{ \bm{W}}_{op}\widebar{\bm{V}}_{pp}\widetilde{ \bm{W}}_{op}^\intercal= V_{\orr \orr} + \sum_{l \neq i}(\widetilde{\bm{W}}_{ol})^2V_{ll}
\end{equation}

Fixed the intervention variable $\ir$, for each outcome variable $\orr$ in $\{1, \dots, n\}\setminus \{\ir\}$, 
\begin{equation}
    \widetilde{\bm{W}}_{\orr l}= (\bm{I}-\widetilde{\bm{B}})^{-1}_{\orr l}\quad \text{for all } l \neq \ir
\end{equation}
where $\widetilde{\bm{B}}$ is obtained from matrix $\bm{B}$ discarding the effects of the intervention variable, i.e. \begin{equation}
    \widetilde{\bm{B}}_{hj}= \begin{cases}
            0 \quad &\text{if } \, j=\ir \\
            \bm{B}_{hl} \quad &\text{otherwise}.
            \end{cases}
\end{equation}
Moreover, calling $\widetilde{\bm{V}}=\text{diag}(v_{1}, \dots, v_{\ir -1}, 0, v_{\ir +1}, \dots, v_n)$, the variances of the post-intervention distribution $\Yor \mid \Do(\Yir = \yir)$ are given by the diagonal terms of 
\begin{equation}
    \widetilde{\bm{\Sigma}} = (\bm{I}-\widetilde{\bm{B}})^{-1}\widetilde{\bm{V}}(\bm{I}-\widetilde{\bm{B}})^{-\intercal} .
\end{equation}
\end{proof}
\subsection{Proof of  \texorpdfstring{\cref{GenDistribution}}{GenDistribution}}\label{ProofTh5}
\begin{proof}
Let $\phi(\cdot)$ and $\Phi(\cdot)$ be respectively the p.d.f. and c.d.f. on a standard normal distribution. By \cref{disgeninter}, $\yor \mid \Do(\Yir = \tilde \yir) \sim \mathcal{N}(\widetilde{\bm{\mu}}_\orr, \widetilde{\bm{\Sigma}}_{\orr \orr})$ therefore, its standardization  $$z_\orr = \frac{\yor- \widetilde{\bm{\mu}}_\orr}{\sqrt{\widetilde{\bm{\Sigma}}_{\orr \orr}}} \sim \mathcal{N}(0,1).$$
Let $\Yir \sim \mathcal{N}(\mu_\ir, \sigma_\ir^2)$ be the marginal distribution of the intervention variable, with $\sigma_\ir = \sqrt{\bm{\Sigma}_{\ir \ir}}$, and $$z_\ir = \frac{\tilde \yir - \mu_\ir}{\sigma_\ir} \sim \mathcal{N}(0,1)$$ its standardization.

Call $$\widebar{\alpha}(o,k) = \frac{\alpha(o,k)-\widetilde{\bm{\mu}}_\orr}{\sqrt{\widetilde{\bm{\Sigma}}_{\orr \orr}}}=\frac{\alpha(o,k)-\mu_\orr}{\sqrt{\widetilde{\bm{\Sigma}}_{\orr \orr}}} - \frac{\bm{W}_{oi}\sigma_\ir}{\sqrt{\widetilde{\bm{\Sigma}}_{\orr \orr}}}z_\ir = a_k+bz_\ir \quad \text{for all } 1 \leq k \leq L_\orr$$ the standardized outcome thresholds and 
$$\widebar{\alpha}(\ir,l) = \frac{\alpha(\ir,l)-\mu_\ir}{\sigma_\ir} \quad \text{for all } 1 \leq l \leq L_\ir$$ the standardized intervention thresholds.
Lastly, let $$\tilde{a}_k=\frac{a_k}{\sqrt{1+b^2}}, \quad \rho = -\frac{b}{\sqrt{1+b^2}}.$$
Then we have
    \begin{align*}
    &\Prob\Big[\Yor \in [\alpha(o,k-1),\alpha(o,k)] \Big]\mid_{f^{**}(\yir')} -\Prob\Big[\Yor \in [\alpha(o,k-1),\alpha(o,k)]\Big] \mid_{f^*(\tilde \yir)}
        \\
        &=  
        \int_{\alpha(o,k-1)}^{\alpha(o,k)}\int_{\alpha(\ir,l'-1)}^{\alpha(1,l')}d\mathcal{N}(\yor,\widetilde{\bm{\mu}}_\orr, \widetilde{\bm{\Sigma}}_{\orr \orr})f^{**}(\yir')\text{d}\yir'\text{d}\yor 
        -\int_{\alpha(o,k-1)}^{\alpha(o,k)}\int_{\alpha(\ir,l-1)}^{\alpha(\ir,l)} d\mathcal{N}(\yor,\widetilde{\bm{\mu}}_\orr, \widetilde{\bm{\Sigma}}_{\orr \orr})f^{*}(\tilde \yir)\text{d}\tilde \yir\text{d}\yor\\
        & \overset{(*)}{=}  \int_{\alpha(\ir,l'-1)}^{\alpha(1,l')} \int_{\widebar{\alpha}(o,k-1)}^{\widebar{\alpha}(o,k)} \phi(z_\orr) \text{d}z_\orr \,  f^{**}(\yir')\text{d}\yir' \, - \int_{\alpha(\ir,l-1)}^{\alpha(\ir,l)} \int_{\widebar{\alpha}_(o,k-1)}^{\widebar{\alpha}(o,k)} \phi (z_\orr)\text{d}z_\orr  \,f^{*}(\tilde \yir)\text{d}\tilde \yir\\
        &= \int_{\alpha(\ir,l'-1)}^{\alpha(1,l')} \Big [\Phi(\widebar{\alpha}(o,k))-\Phi(\widebar{\alpha}(o,k-1))\Big]  \,  \frac{1}{\sigma_\ir} \frac{\phi(\frac{\yir'-\mu_\ir}{\sigma_\ir})}{\Phi(\frac{\alpha(\ir,l')-\mu_\ir}{\sigma_\ir})- \Phi(\frac{\alpha(\ir,l'-1)-\mu_\ir}{\sigma_\ir})}\text{d}\yir' \, \\
        & \quad - \int_{\alpha(\ir,l-1)}^{\alpha(\ir,l)} \Big [\Phi(\widebar{\alpha}(o,k))-\Phi(\widebar{\alpha}(o,k-1))\Big] \, \frac{1}{\sigma_\ir} \frac{\phi(\frac{\tilde \yir-\mu_\ir}{\sigma_\ir})}{\Phi(\frac{\alpha(\ir,l)-\mu_\ir}{\sigma_\ir})- \Phi(\frac{\alpha(\ir,l-1)-\mu_\ir}{\sigma_\ir})}\text{d}\tilde \yir \\
        &= \int_{\widebar{\alpha}(\ir,l'-1)}^{\widebar{\alpha}(\ir,l')} \Big [\Phi(a_k+bz'_\ir)-\Phi(a_{k-1}+bz'_\ir)\Big] \,  \frac{\phi(z'_\ir)}{\Phi(\widebar{\alpha}(\ir,l'))- \Phi(\widebar{\alpha}(\ir,l'-1)}\text{d}z'_\ir \, \\ 
        & \quad - \int_{\widebar{\alpha}(\ir,l-1)}^{\widebar{\alpha}(\ir,l)} \Big [\Phi(a_k+bz_\ir)-\Phi(a_{k-1}+bz_\ir)\Big] \,  \frac{\phi(z_\ir)}{\Phi(\widebar{\alpha}(\ir,l))- \Phi(\widebar{\alpha}(\ir,l-1)}\text{d}z_\ir\,\\
        &\overset{(**)}{=}
\frac{1}{\Phi(\widebar{\alpha}(\ir,l'))- \Phi(\widebar{\alpha}(\ir,l'-1)}
\quad \begin{aligned}
  \Big[ &\mathcal{BN}\left(\widebar{\alpha}(\ir,l'), \tilde{a}_k, \rho\right)  
  - \mathcal{BN}\left(\widebar{\alpha}(\ir,l'-1), \tilde{a}_k, \rho\right) \\
   & -\mathcal{BN}\left(\widebar{\alpha}(\ir,l'), \tilde{a}_{k-1}, \rho\right) 
   + \mathcal{BN}\left(\widebar{\alpha}(\ir,l'-1), \tilde{a}_{k-1}, \rho\right) \Big]
        \end{aligned}\\
& - \frac{1}{\Phi(\widebar{\alpha}(\ir,l))- \Phi(\widebar{\alpha}(\ir,l-1)} 
  \quad \begin{aligned}
  \Big[ &\mathcal{BN}\left(\widebar{\alpha}(\ir,l), \tilde{a}_k, \rho\right) 
  - \mathcal{BN}\left(\widebar{\alpha}(\ir,l-1), \tilde{a}_k, \rho\right) \\
  &- \mathcal{BN}\left(\widebar{\alpha}(\ir,l), \tilde{a}_{k-1}, \rho\right) 
  + \mathcal{BN}\left(\widebar{\alpha}(\ir,l-1), \tilde{a}_{k-1}, \rho\right)\Big].
        \end{aligned}\label{OCEtheorem2}
        \end{align*}
            
where we use Fubini-Tonelli theorem to exchange the order of integration in $(*)$ and the following formulas from \citet{owenTableNormalIntegrals1980}, where the numbers in parentheses referred to the formula's number in the paper:    
\begin{equation*}
\begin{split}
        \int_h^k \phi(x)\Phi (a+bx)dx&=\int_{-\infty}^{\frac{a}{\sqrt{b^2+1}}} \phi(x)\Phi (k\sqrt{b^2+1}+bx)dx \\
        &-\int_{-\infty}^{\frac{a}{\sqrt{b^2+1}}} \phi(x)\Phi (h\sqrt{b^2+1}+bx)dx  \quad (10,010.4)  \\
        \int_{-\infty}^Y \Phi (a+bx)\phi(x)dx &= \mathcal{BN}\Bigg [ \frac{a}{\sqrt{1+b^2}},Y;\rho= \frac{-b}{\sqrt{1+b^2}}\Bigg] \quad  (10,010.1) \\
        \end{split}
\end{equation*}
\begin{equation*}
\begin{split}
     \mathcal{BN}(h,k;\rho) &= \begin{cases}
            \frac{1}{2}\Phi (h)-T\Big(h,\frac{k-\rho h}{h\sqrt{1-\rho^2}}\Big)+\frac{1}{2}\Phi (k) - T\Big(k,\frac{h-\rho k}{k\sqrt{1-\rho^2}}\Big) \\
            \text{if}\, hk>0, \, \text{or if }\, hk=0 \,\text{and}\, h \,\text{or}\, k >0, \,\text{or if both } = 0 \\
            \frac{1}{2}\Phi (h)-T\Big(h,\frac{k-\rho h}{h\sqrt{1-\rho^2}}\Big)+\frac{1}{2}\Phi (k) - T\Big(k,\frac{h-\rho k}{k\sqrt{1-\rho^2}}\Big)-\frac{1}{2} \\
            \text{if}\, hk<0, \,\text{or if}\, hk=0 \, \text{and}\, h \,\text{or}\, k>0.
        \end{cases} \quad (3.1)
        \end{split}
\end{equation*}
\end{proof} 
\subsection{Equivalence of Distribution and Quantile Approached for Truncated Normal Intervention Policies}\label{sup-equiv}
Although the proof is demonstrated for the toy model in \cref{apptoymodel}, it can be easily extended to more general cases involving additional variables with multiple ordinal levels.  

\begin{equation}
    \begin{split}
    \Prob\Big[Y_2 \in &[\alpha(2,k-1),\alpha(2,k)]\Big]|_{f^{**}(y_1')} - \Prob\Big[Y_2 \in [\alpha(2,k-1),\alpha(2,k)]\Big]|_{f^{*}(\tilde y_1)} -  \\ 
    &=\int_{\alpha(2,k-1)}^{\alpha(2,k)}\int_{\alpha(1,l'-1)}^{\alpha(1,l')}\frac{1}{\sqrt{2\pi v_2^2}}\exp\Big \{\frac{1}{2v_2^2}\big(y_2 - \mu_2 - b_{12}(y_1'-\mu_1)\big)^2\Big \}f^{**}(y_1')\text{d}y_1'\text{d}y_2\\
    &\quad - \int_{\alpha(2,k-1)}^{\alpha(2,k)}\int_{\alpha(1,l-1)}^{\alpha(1,l)} \frac{1}{\sqrt{2\pi v_2^2}} \exp\Big \{\frac{1}{2v_2^2}\big(y_2 - \mu_2 - b_{12}(\tilde y_1-\mu_1)\big)^2\Big \}f^{*}(\tilde y_1)\text{d}\tilde y_1\text{d}y_2\\
    \intertext{(distributional approach)}
    & \underset{\cref{changevar}}{=}\int_{\alpha(2,k-1)}^{\alpha(2,k)}\int_{\alpha(1,l-1)}^{\alpha(1,l)} \frac{1}{\sqrt{2\pi v_2^2}} \Bigg [\exp\Big \{\frac{1}{2v_2^2}\big(y_2 - \mu_2 - b_{12}(F^{{**}^{-1}}(F^{*}(\tilde y_1))-\mu_1)\big)^2\Big \} \\
    &\quad - \exp\Big \{\frac{1}{2v_2^2}\big(y_2 - \mu_2 - b_{12}(\tilde y_1-\mu_1)\big)^2\Big \}\Bigg ]f^{*}(\tilde y_1)\text{d}\tilde y_1\text{d}y_2 \\
    \intertext{(quantile approach).}
    \end{split}
\end{equation}

\section{Additional Experimental Results}
For the simulations and the analysis we used the \textsf{R} statistical software \citep[v.4.4.1]{statsr}, where we carried out alternative implementations of the formula in \cref{GenDistribution}, the first using the numerical integration \textsf{integrate} function from the stats package, the second exploiting the bivariate normal distribution \textsf{pnorm} function from the \textsf{mvtnorm} package \citep[v.1.1-3]{mvtnorm}, and the last using Owen’s T \textsf{OwenT} function from the \textsf{OwenQ} package \citep[v.1.0.7]{OwenQ} for the computations of integrals. Because, as expected, all implementations delivered equivalent results (up to numerical errors) as shown for both the Simulation and the Application to Psychological Data in the Supplementary Material. We present here the results delivered by numerical integration. 
\subsection{The binary case}\label{AP_Bin}

In a simplified scenario where we only have two binary variables, without any hidden confounding, we can estimate the average causal effect of the parent node on the child node directly from a contingency table.

Consider two binary variables $X_1$ and $X_2$, whose relationship is described by the following contingency table and $X_1$ is a direct cause of $X_2$. 

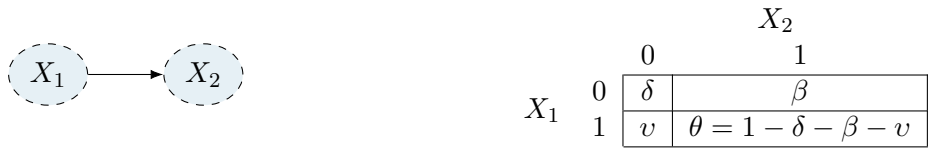
\begin{figure}[ht]
    \centering
    \begin{minipage}[]{0.45\textwidth}
        \centering
        \begin{tikzpicture}[scale=0.7,baseline=(current bounding box.north), state/.style={circle, draw=blue!30!black, fill=blue!30, minimum size=2em}]
            \node[da, fill=blue!60!green!10] (j)  {$X_1$};
            \node[da, fill=blue!60!green!10] (k) [right =of j] {$X_2$};
            \path (j) edge (k);
        \end{tikzpicture}
    \end{minipage}%
    \hfill
    \begin{minipage}[]{0.5\textwidth}
        \centering
   \begin{tabular}{cc|c|c|}
        
        & \multicolumn{1}{c}{} & \multicolumn{2}{c}{$X_2$}\\
        & \multicolumn{1}{c}{} & \multicolumn{1}{c}{$0$}  & \multicolumn{1}{c}{$1$} \\\cline{3-4}
        \multirow{2}*{$X_1$}  & $0$ & $\delta$ & $\beta$ \\\cline{3-4}
        & $1$ & $\upsilon$ & $\theta =1-\delta-\beta-\upsilon$ \\\cline{3-4}
        \end{tabular}
    \end{minipage}
    \caption{Two binary variable case: DAG (left) and the corresponding contingency table (right).}
    \label{twovarDAG_table}
\end{figure}

On the scale of the risk difference, the average causal effects of $X_1$ on $X_2$, are determined as
\begin{equation}\label{2varrisk}
    \begin{split}
        I_0 = \Prob[X_2 = 0 \mid &\Do(X_1 = 1)] - \Prob[X_2 = 0 \mid \Do(X_1 = 0)] = \frac{\upsilon}{\upsilon + \theta}-\frac{\delta}{\delta+\beta}\\
        I_1 = \Prob[X_2 = 1 \mid &\Do(X_1 = 1)] - \Prob[X_2 = 1 \mid \Do(X_1 = 0)] =  \frac{\theta}{\upsilon + \theta}-\frac{\beta}{\delta + \beta}.  
    \end{split}
\end{equation}

Alternatively, we may model the relationship between $X_1$ and $X_2$ as a latent Gaussian DAG model as defined in \cref{seclatentmodel}. Because 3 independent parameters are sufficient to describe the joint distribution of $X_1$ and $X_2$, a latent Gaussian DAG model can also fully characterise it in terms of the covariance between $X_1$ and $X_2$ and the two thresholds that discretise, for each variable, the underlying continuous Gaussian distribution into the two corresponding ordinal levels. Therefore, if our latent model is coherent, we expect it to deliver the same results as those obtained directly from the contingency table. In the following, we verify that assuming a Latent Gaussian DAG model comprising only the intervention and outcome variables (as in the toy model in \cref{apptoymodel}) indeed yields the same results.
We can express the causal effect on the two levels of $X_2$ due to an intervention shifting $X_1$ from level 0 to level 1 as following, adopting the latent construction described in \cref{toymodeleffect}.    
\begin{equation}\label{risk0}
        \widetilde{I_0} = \Prob\Big[Y_2 \in [-\infty,\alpha_2]\mid \Do(Y_1 \in [\alpha_1,+\infty])\Big]
        \ -  \Prob\Big[Y_2 \in [-\infty,\alpha_2] \mid \Do(Y_1 \in [-\infty, \alpha_1])\Big]
        \end{equation}
        \begin{equation}\label{risk1}
        \widetilde{I_1} = \Prob\Big[Y_2 \in [\alpha_2, \infty]\mid \Do(Y_1 \in [\alpha_1,+\infty])\Big]
        \ -  \Prob\Big[Y_2 \in [\alpha_2, \infty] \mid \Do(Y_1 \in [-\infty, \alpha_1])\Big], 
\end{equation}
where $\boldsymbol{Y} = (Y_1, Y_2)^\intercal \sim \mathcal{N}(\boldsymbol{0}, \boldsymbol{\Sigma})$, $b$ is the given regression coefficient of $Y_2$ on $Y_1$, $\bm{\alpha}=(\alpha_1, \alpha_2)$ the given thresholds. 

Setting $b = 0.5$, $\bm{\alpha}= (0.2,0.4)$ and $\bm{V} = \bm{I}$ one can compute \cref{risk0} and \cref{risk1} with the previously defined distribution and quantile strategies. We denote these respectively as $(\widetilde{I}_{0d}, \widetilde{I}_{1d})$ and $(\widetilde{I}_{0q}, \widetilde{I}_{1q})$.

\begin{table}[!h]
\centering
{\tabcolsep=4.25pt
\begin{tabular}{@{}cccccc@{}}
$I_0$ &   $\widetilde{I}_{0d}$ & $\widetilde{I}_{0q}$ & $I_1$ &   $\widetilde{I}_{1d}$ & $\widetilde{I}_{1q}$ \\
\hline
  -0.281642 & -0.2816425 & -0.2816404 & 0.281642 & 0.281642 & 0.2816419 \\
\end{tabular}}
\caption{Results of causal effect estimation for the binary case. 
\label{Table1}}
\end{table}

As shown in \cref{Table1}, for two binary variables, the proposed method for computing causal effects determines results in agreement with the causal risk differences. However, this equivalence no longer holds for three or more binary variables. While the latent Gaussian construction can fully characterize the joint distribution of two binary variables, it is well known that no Gaussian distribution can fully represent the unconstrained joint distribution of three or more binary variables. \cref{AP_3Var} illustrates the problem in a case with 3 binary variables. Consequently, the latent Gaussian approach cannot fully capture the distributional properties of ordinal variables, leading to a loss of resolution as dimensionality increases.

Moreover, as proved in \cref{sup-equiv}, causal effects $(\widetilde{I}_{0d}, \widetilde{I}_{1d})$ and $(\widetilde{I}_{0q}, \widetilde{I}_{1q})$ computed respectively through the distribution and quantile strategies are also equivalent, up to numerical precision in the computation of integrals.

\subsection{Three Binary Variables Case }\label{AP_3Var}
Consider the case of three binary ordinal variables, described by the Latent Gaussian DAG and the following probability tables: 

\begin{figure}[ht]
    \centering
    \begin{minipage}[]{0.45\textwidth}
        \centering
        \begin{tikzpicture}[scale=0.7,baseline=(current bounding box.north), state/.style={circle, draw=blue!30!black, fill=blue!30, minimum size=2em}]
            \node[state] (m) at (0,0) {$Y_1$};
            \node[da, fill=blue!60!green!10] (v) [right =of m] {$X_1$};
            \node[state] (j) at (-2,-2) {$Y_2$};
            \node[da, fill=blue!60!green!10] (b) [below =of j] {$X_2$};
            \node[state] (k)  at (2,-2) {$Y_3$};
            \node[da, fill=blue!60!green!10] (c) [below =of k] {$X_3$};
            \path (m) edge (j);
             \draw (m) -- (j) node [midway,sloped, fill=white] {$b_{12}$};
            \draw[dashed,draw=blue!30!black, ->] (m) -- (v);
            \draw[dashed,draw=blue!30!black, ->] (j) -- (b);
            \draw[dashed,draw=blue!30!black, ->] (k) -- (c);
            \path (m) edge (k);
             \draw (m) -- (k) node [midway, sloped, fill=white] {$b_{13}$};
            \path (j) edge (k);
             \draw (j) -- (k) node [midway, fill=white] {$b_{23}$};
        \end{tikzpicture}
    \end{minipage}%
    \hfill
    \begin{minipage}[]{0.5\textwidth}
        \centering
        \begin{tabular}{ c| c | c | l}
            $X_1$ &  $X_2$ & $X_3$ & $p$ \\
            \hline
            0 &  0 & 0 & $\tau_0$ \\ 
            0 &  0 & 1 & $\tau_1$ \\ 
            0 &  1 & 0 & $\beta_0$ \\ 
            0 &  1 & 1 & $\beta_1$ \\ 
            1 &  0 & 0 & $\gamma_0$ \\ 
            1 &  0 & 1 & $\gamma_1$ \\ 
            1 &  1 & 0 & $\theta_0$ \\ 
            1 &  1 & 1 & $\theta_1$ \\ 
        \end{tabular}
    \end{minipage}
    \caption{Three variable case: DAG (left) and the corresponding contingency table (right).}
    \label{threevarDAG_table}
\end{figure}
From the previous contingency tables, one can derive: 
\begin{equation}
    \begin{split}
        &p_{X_1} =\Prob(X_1 = 1) = \gamma_0+\theta_0+\theta_1+\gamma_1, \\
        & p_{X_2,0} =\Prob(X_2 = 1 \mid X_1 = 0) =  \frac{\beta_0+\beta_1}{1-p_{X_1}}, \\
        & p_{X_2,1}=\Prob(X_2 = 1 \mid X_1 = 1) = \frac{\theta_0+\theta_1}{p_{X_1}}  \\
        & p_{X_3,0}=\Prob(X_3 = 1 \mid X_1 = 0, X_2 = 0) = \frac{\tau_1}{\tau_0+\tau_1}, \\  
        &p_{X_3,1}=\Prob(X_3 = 1 \mid X_1 = 0, X_2 = 1) = \frac{\beta_1}{\beta_0+\beta_1}, \\ &p_{X_3,2}=\Prob(X_3 = 1 \mid X_1 =1, X_2 = 0) = \frac{\gamma_1}{\gamma_0+\gamma_1} ,\\
        &p_{X_3,3}=\Prob(X_3 = 1 \mid X_1 = 1, X_2 = 1) = \frac{\theta_1}{\theta_0+\theta_1}.
    \end{split}
\end{equation}

In analogy with \cref{2varrisk}, one can compute the following risk differences 
 \begin{equation}\label{risk3}
    \begin{split}
        I_0 &= \Prob[X_3 = 0 \mid \Do(X_2 = 1)] - \Prob[X_3 = 0 \mid \Do(X_2 = 0)] \\ &= [(1-p_{X_3,1})\cdot (1-p_{X_1}) + (1-p_{X_3,3})\cdot p_{X_1}] - [(1-p_{X_3,0})\cdot (1-p_{X_1})+(1-p_{X_3,2})\cdot p_{X_1}] \\
        \\
        I_1 &= \Prob[X_3 = 1 \mid \Do(X_2 = 1)] - \Prob[X_3 = 1 \mid \Do(X_2 = 0)]\\ &= [p_{X_3,1}\cdot (1-p_{X_1}) + p_{X_3,3}\cdot p_{X_1}] - [p_{X_3,0}\cdot (1-p_{X_1})+p_{X_3,2}\cdot p_{X_1}] 
    \end{split}
\end{equation}
and compare with the corresponding differences in latent probabilities obtained from the Gaussian Latent DAG model \begin{equation}\label{risk03}
        \widetilde{I_0} = \Prob\Big[Y_3 \in [-\infty,\alpha_3]\mid \Do(Y_2\in [\alpha_2,+\infty])\Big]
        \ -  \Prob\Big[Y_3 \in [-\infty,\alpha_3] \mid \Do(Y_2 \in [-\infty, \alpha_2])\Big]
        \end{equation}
        \begin{equation}\label{risk13}
        \widetilde{I_1} = \Prob\Big[Y_3 \in [\alpha_3, \infty]\mid \Do(Y_2 \in [\alpha_2,+\infty])\Big]
        \ -  \Prob\Big[Y_3 \in [\alpha_3, \infty] \mid \Do(Y_2 \in [-\infty, \alpha_2])\Big]. 
\end{equation}

Setting $\bm{\mu} = \bm{0}$, $\bm{b}=(b_{12},b_{13}, b_{23})=(0.5,0.8,0.9)$, $\bm{\alpha}=(1.2,2.4,3.3)$ and $\bm{v}=(1,1,1)$ one gets:
\begin{table}[ht]
    \centering
    \begin{tabular}{c|c | c| c}
    \hline
         $I_0$ & -0.3617032 & $I_1$ & 0.3617032 \\
         \hline
         $\widetilde{I}_{0d}$ & -0.2590655 &$\widetilde{I}_{1d}$ & 0.2590634 \\
         $\widetilde{I}_{0q}$ & -0.2590577&$\widetilde{I}_{1q}$ & 0.259063\\
         \hline 
    \end{tabular}
    \caption{Results of causal effect estimation for the three binary variables case. }
    \label{tab:3varres}
\end{table}
As expected, contrary to the previous binary case, it happens that latent probabilities differences in \cref{risk03} and \cref{risk13} do not match the risk differences in \cref{risk3}. In fact, it is well known that no Gaussian distribution can fully characterize an unconstrained distribution of three binary variables.

\subsection{Simulations}\label{AP_Simulation}
\vspace{\fill}

\begin{figure}[H]
    \centering
    \includegraphics[scale=0.4]{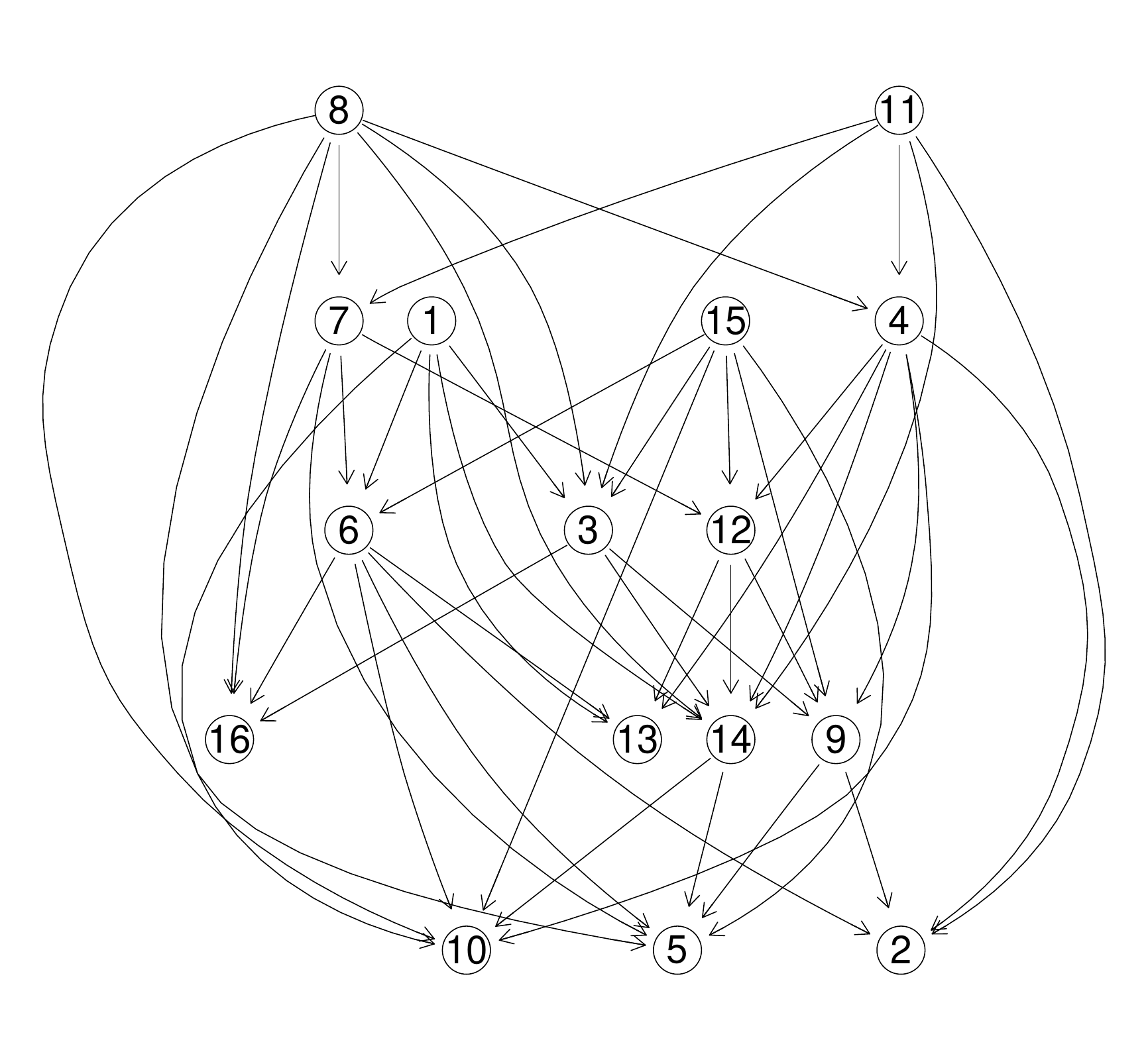}
    \caption{Random E-R DAG with 16 nodes adopted as Latent Gaussian DAG for the simulations. }
    \label{ERGraph}
\end{figure}
\vspace{\fill}

\newpage

\begin{figure}[H]  % H forces images to stay on the page
    \centering
    \begin{minipage}{\textwidth}
        \centering
       \includegraphics[height=0.45\textheight, keepaspectratio]{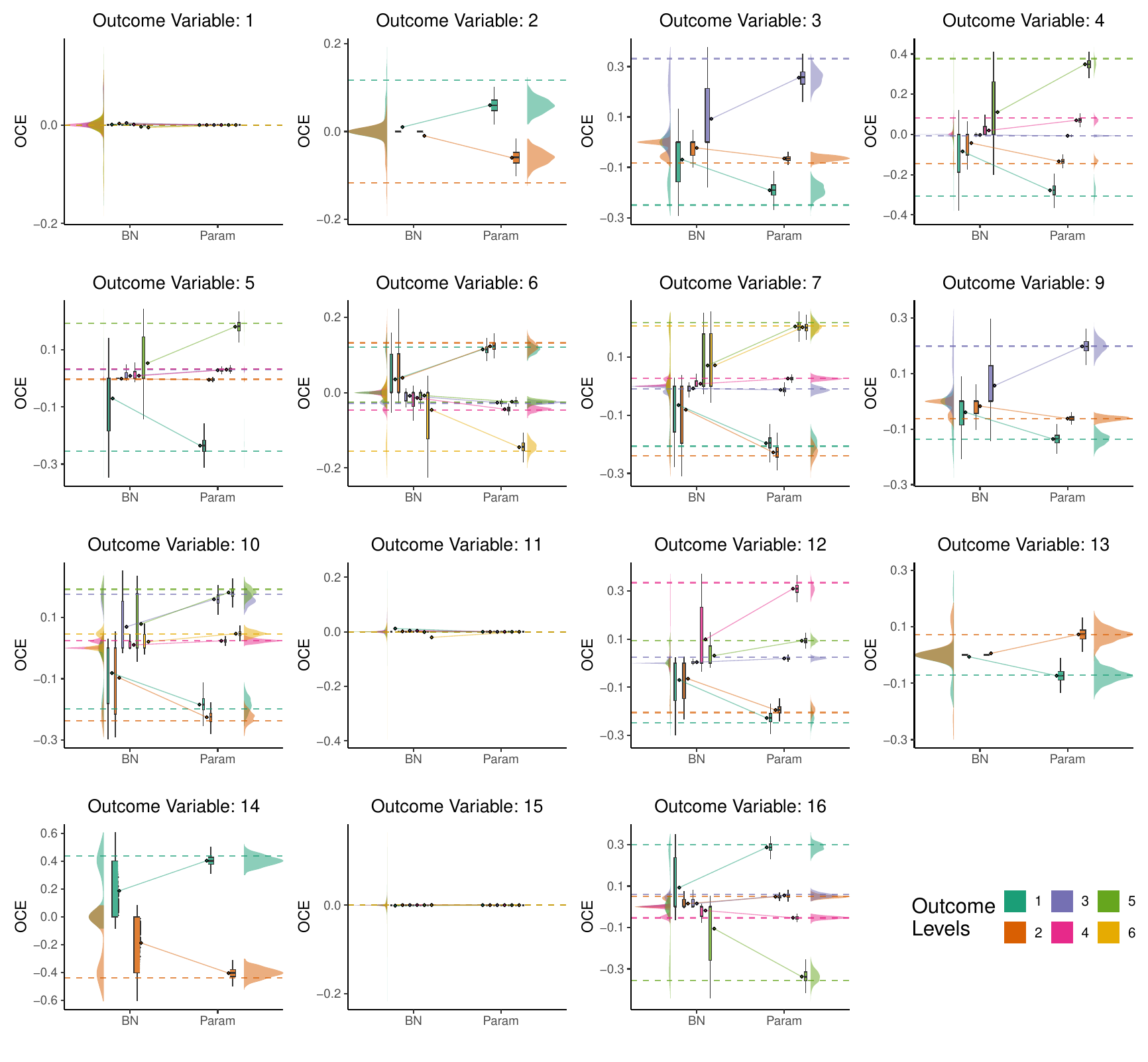}
    \caption{Simulation with Regenerated Data: Ordinal Causal Effects of intervention variable 8.}
    \label{matrixfig8R}
    \end{minipage}
    
    \vspace{0.1cm} % Space between images

    \begin{minipage}{\textwidth}
        \centering
        \includegraphics[height=0.45\textheight, keepaspectratio]{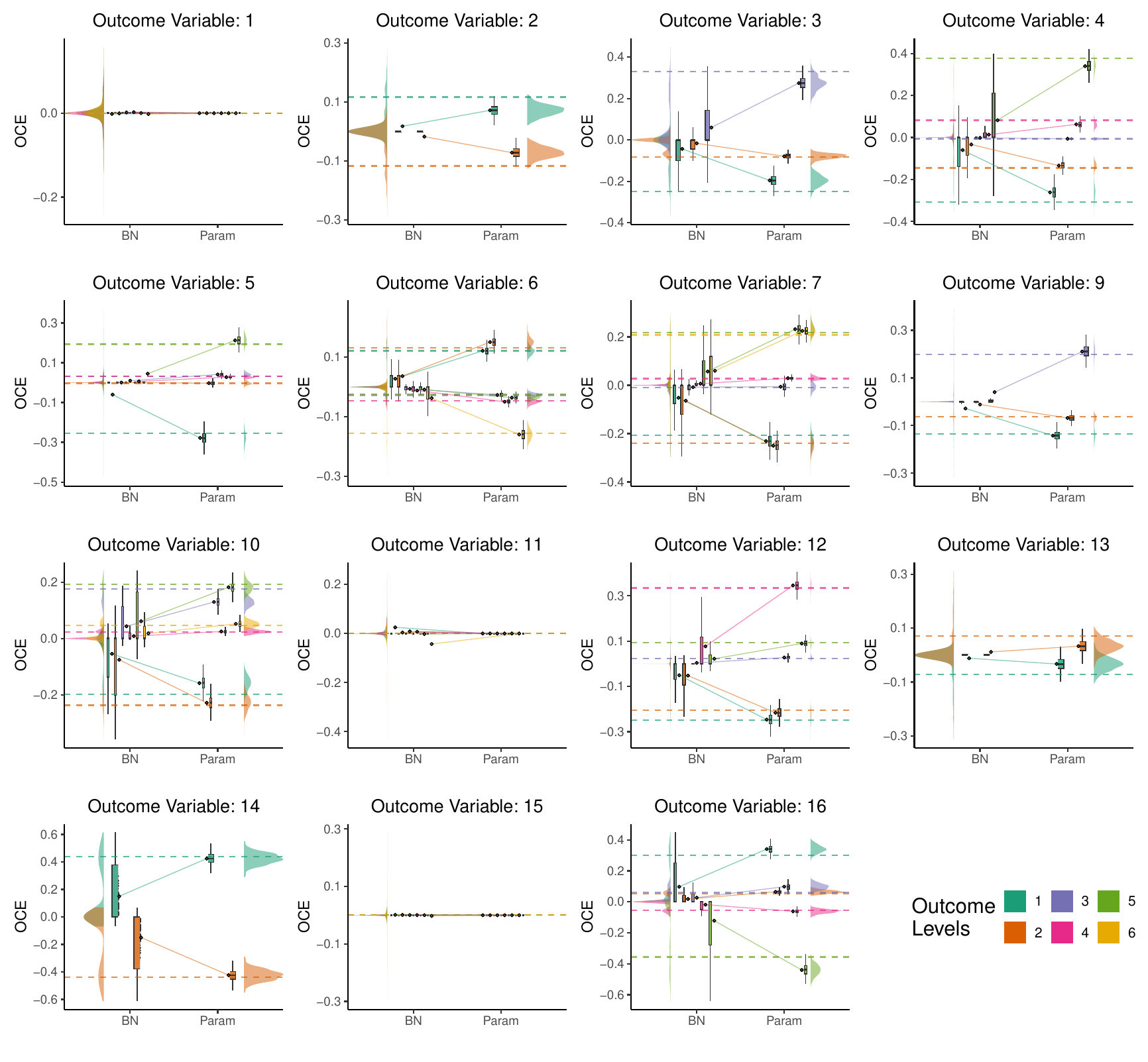}
    \caption{Simulation with Bootstrapped Data: Ordinal Causal Effects of intervention variable 8.}
    \label{matrixfig8B}
    \end{minipage}
\end{figure}

\begin{figure}[H]  % H forces images to stay on the page
    \centering
    \begin{minipage}{\textwidth}
        \centering
       \includegraphics[height=0.45\textheight, keepaspectratio]{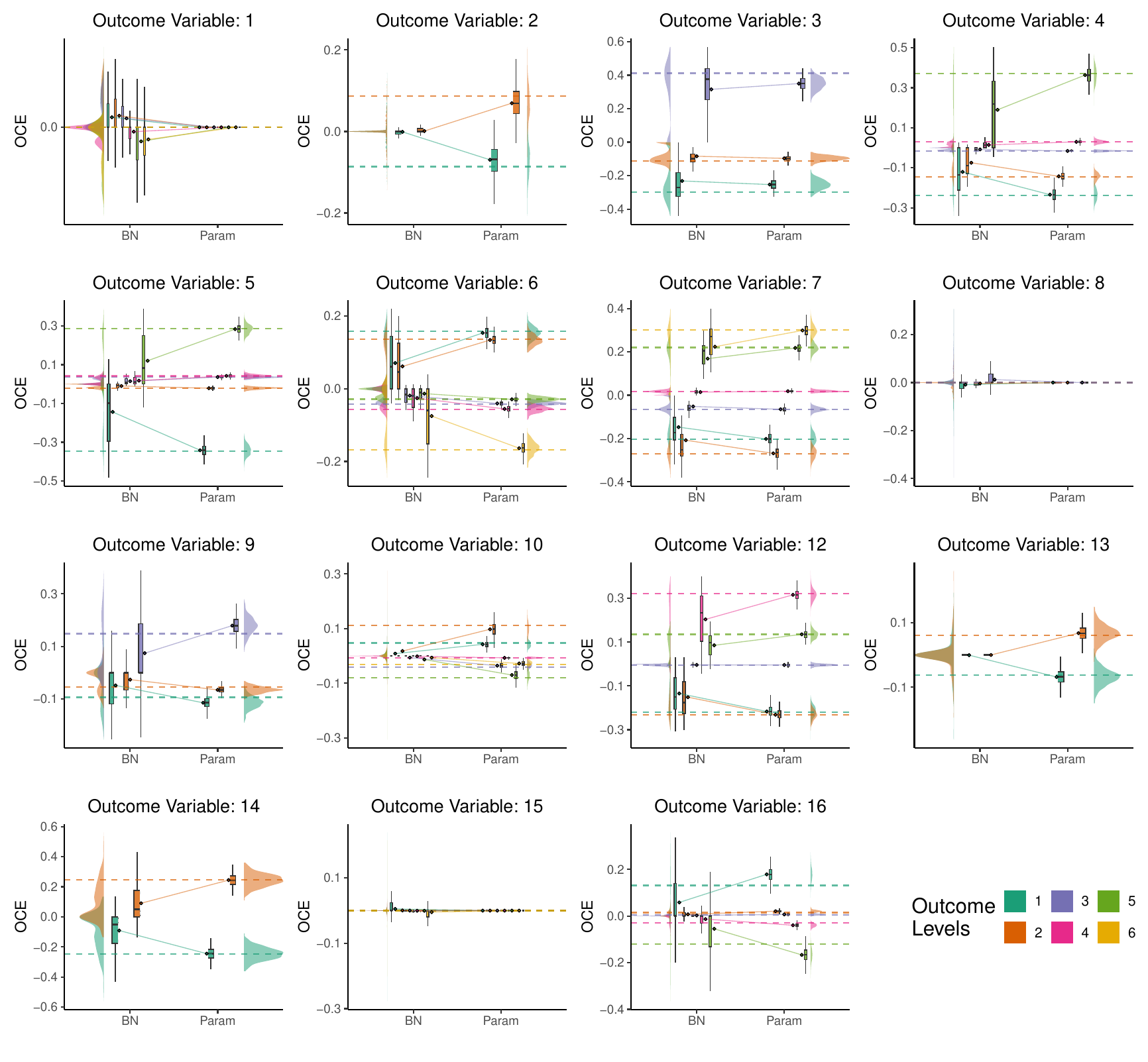}
    \caption{Simulation with Regenerated Data: Ordinal Causal Effects of intervention variable 11.}
    \label{matrixfig11R}
    \end{minipage}
    
    \vspace{0.1cm} % Space between images

    \begin{minipage}{\textwidth}
        \centering
        \includegraphics[height=0.45\textheight, keepaspectratio]{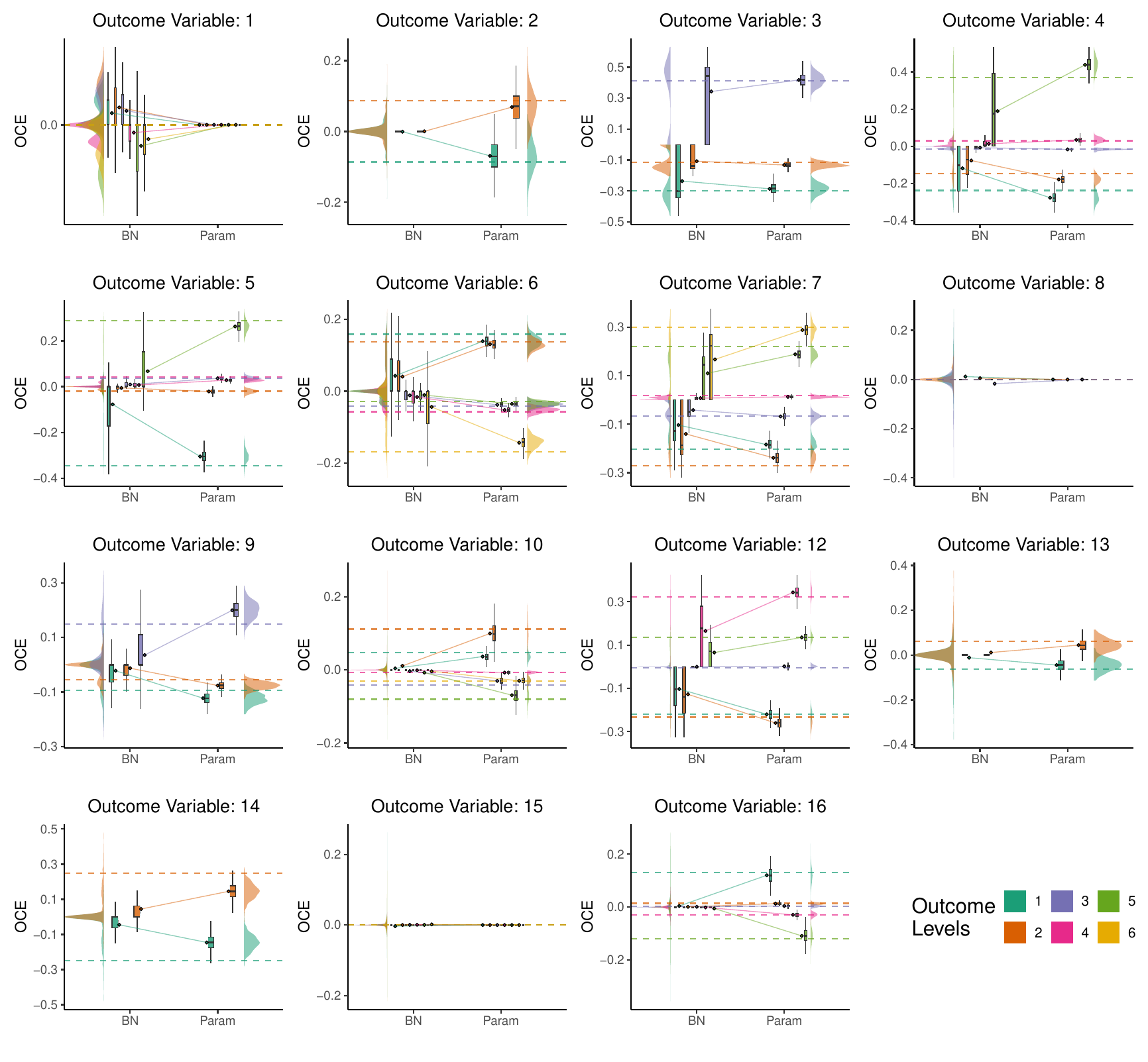}
    \caption{Simulation with Bootstrapped Data: Ordinal Causal Effects of intervention variable 11.}
    \label{matrixfig11B}
    \end{minipage}
\end{figure}

\begin{figure}[H]  % H forces images to stay on the page
    \centering
    \begin{minipage}{\textwidth}
        \centering
       \includegraphics[height=0.45\textheight, keepaspectratio]{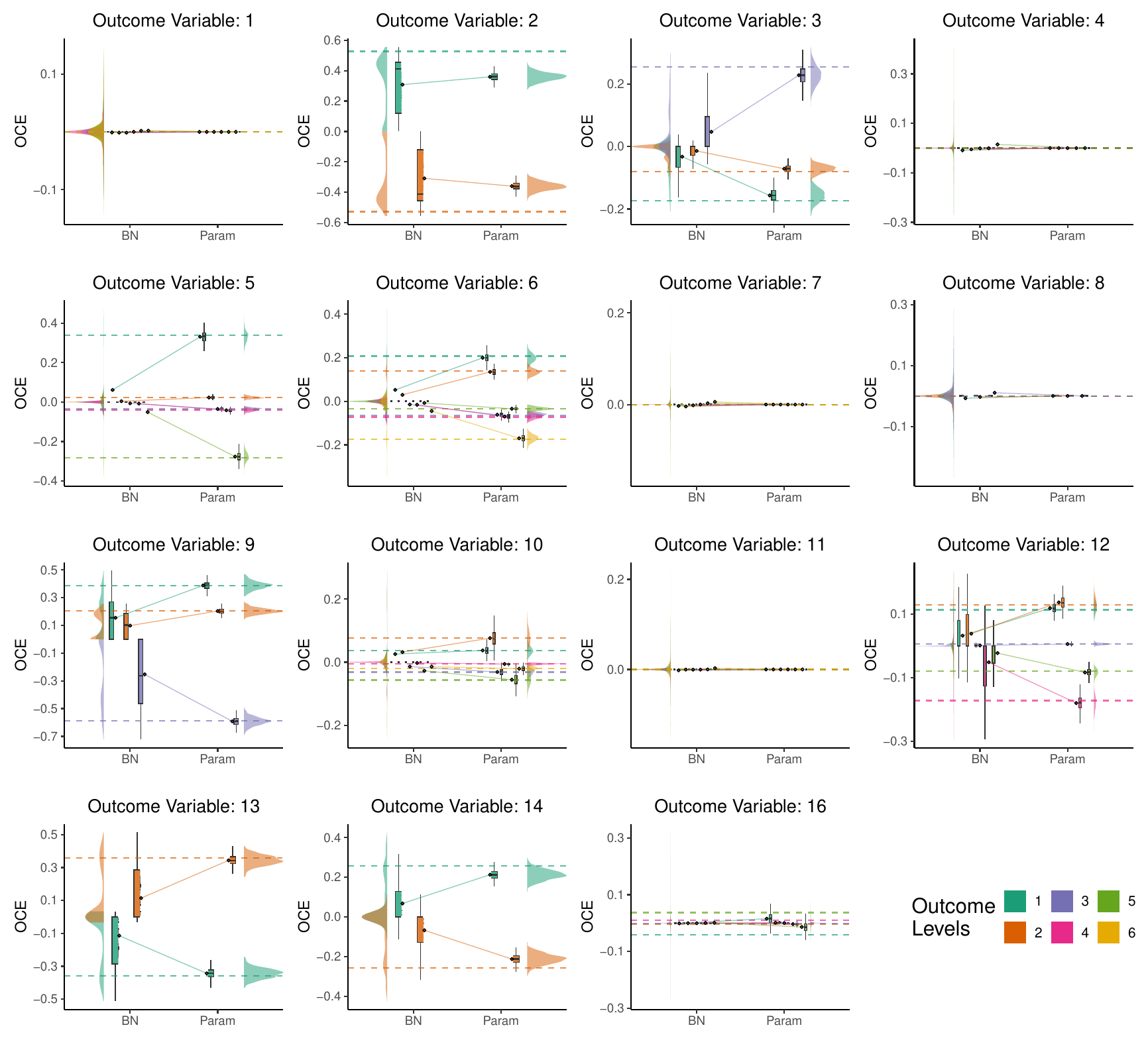}
    \caption{Simulation with Regenerated Data: Ordinal Causal Effects of intervention variable 15.}
    \label{matrixfig15R}
    \end{minipage}
    
    \vspace{0.1cm} % Space between images

    \begin{minipage}{\textwidth}
        \centering
        \includegraphics[height=0.45\textheight, keepaspectratio]{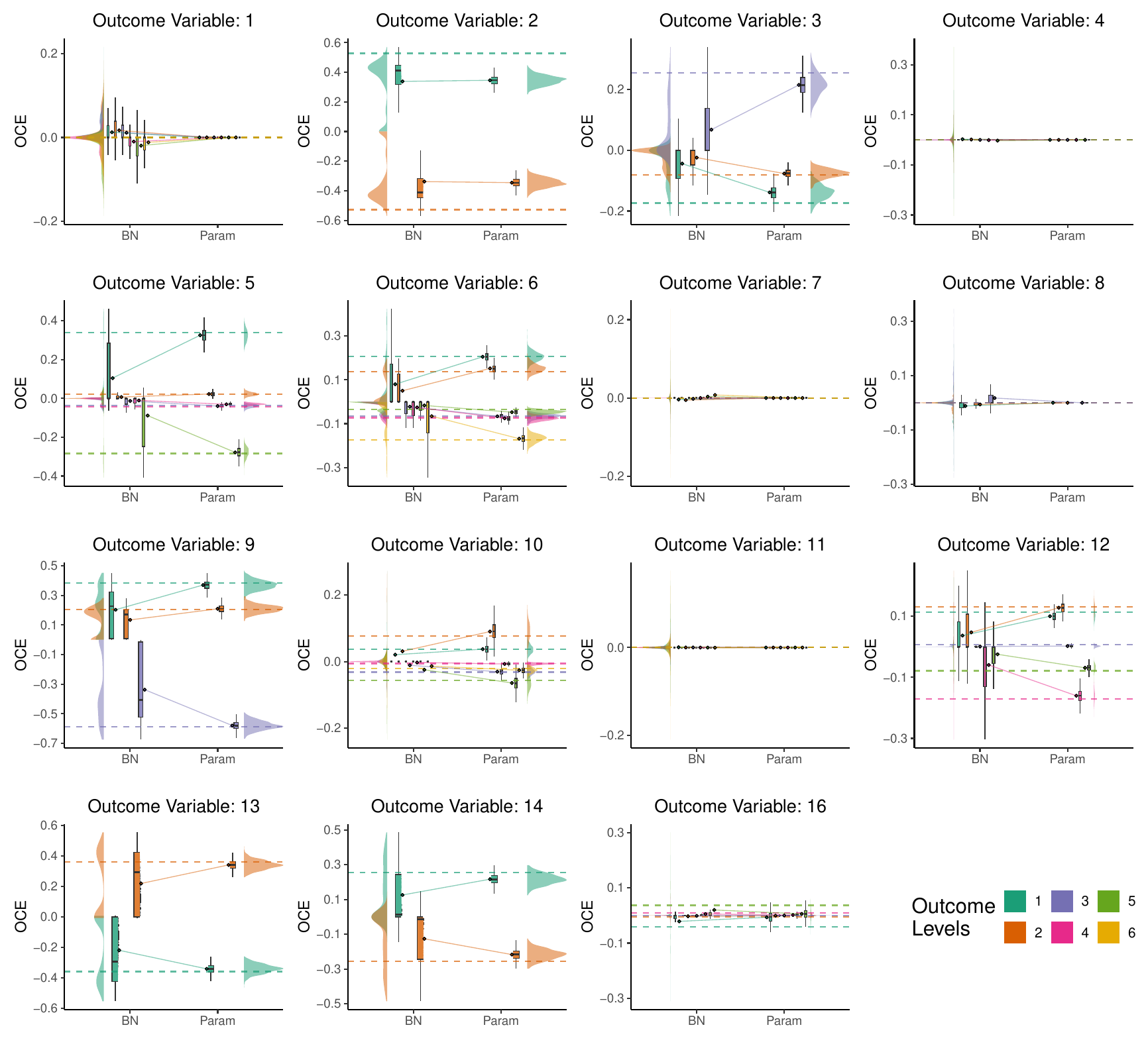}
    \caption{Simulation with Bootstrapped Data: Ordinal Causal Effects of intervention variable 15.}
    \label{matrixfig15B}
    \end{minipage}
\end{figure}

\subsection{Analysis of Pyschological Data}

\vspace{\fill} 

\begin{figure}[h]
    \centering
    \includegraphics[scale=0.6]{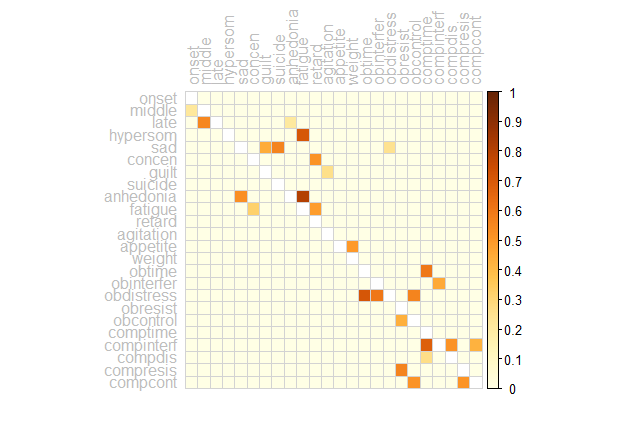}
    \caption{Heatmaps for the CPDAG adjacency matrices of psychological survey data of \citet{McNally}. The darker the shade in the grid, the more frequently the corresponding directed edge occurs in the 500 Bootstrapped CPDAGs; where an indirect edge counts half for each direction. }
    \label{Heatmap}
\end{figure}

\vspace{\fill} 
\newpage 
\begin{figure}[H]
    \centering
\includegraphics[width=15cm, height = 21cm]{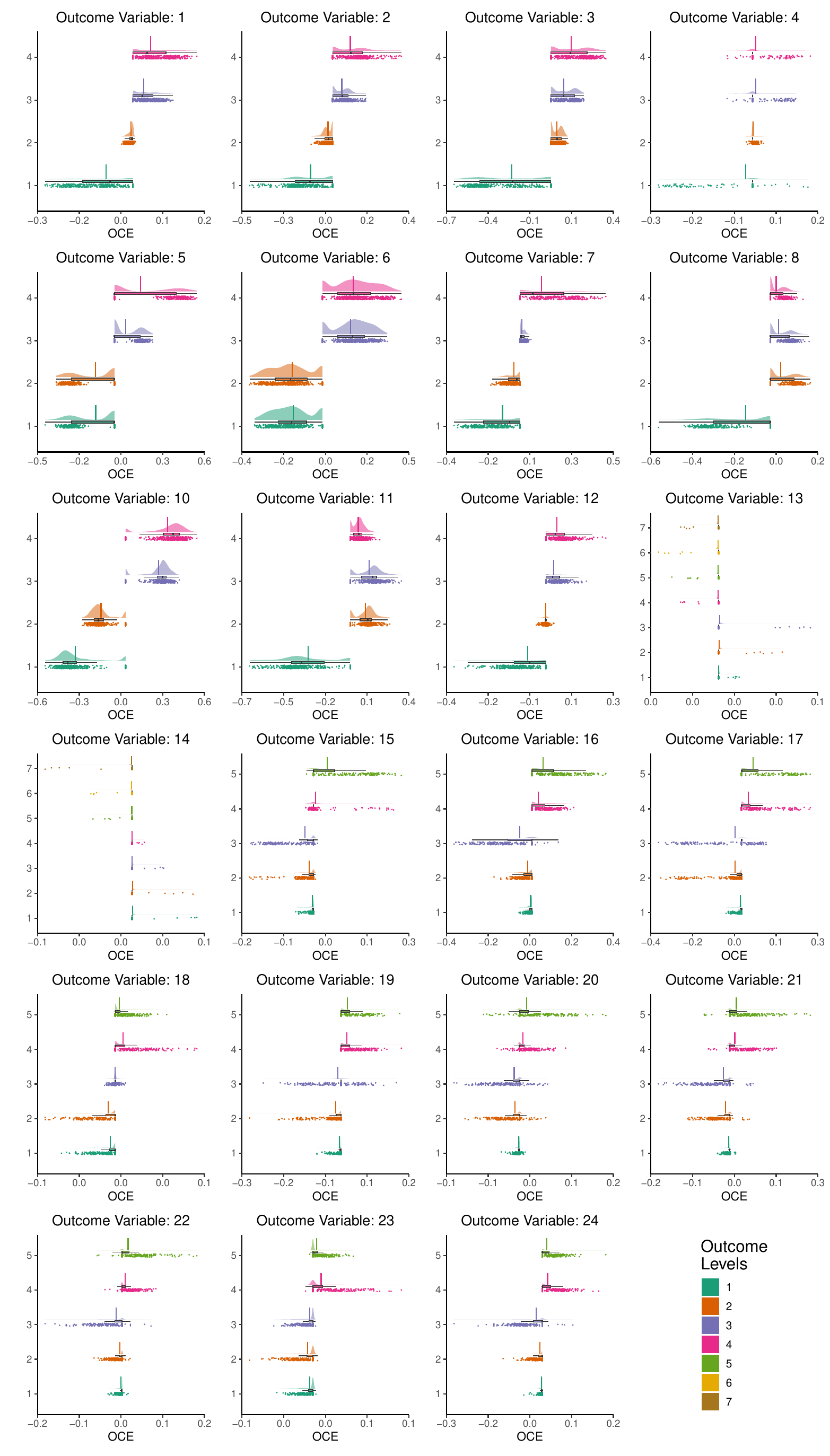}
    \caption{Ordinal Causal Effect of variable 9, resulting from switching it from its lowest to the largest level. The solid lines corresponds to the mean of OCEs for each level of the outcome variable. }
    \label{Effect9}
\end{figure}

\end{document}